\documentstyle [12pt,epsf] {article}

\begin{document}
\begin{titlepage}

\begin{center}
\vspace{2cm}
\LARGE
The Dependence of Star Formation History and Internal Structure on
Stellar Mass for $10^5$  Low-Redshift Galaxies
\\                                                     
\vspace{1cm} 
\large
Guinevere Kauffmann$^1$, Timothy M. Heckman$^2$, Simon D.M. White$^1$,   
St\'ephane Charlot$^{1,3}$, Christy Tremonti$^2$, Eric W. Peng$^2$,
Mark Seibert$^2$, Jon Brinkmann$^4$, Robert C. Nichol$^5$, Mark SubbaRao$^6$,
Don York$^6$ \\
\vspace{0.3cm}
\small
{\em $^1$Max-Planck Institut f\"{u}r Astrophysik, D-85748 Garching, Germany} \\
{\em $^2$Department of Physics and Astronomy, Johns Hopkins University, Baltimore, MD 21218}\\
{\em $^3$ Institut d'Astrophysique du CNRS, 98 bis Boulevard Arago, F-75014 Paris, France} \\
{\em $^4$ Apache Point Observatory, P.O. Box 59, Sunspot, NM 88349} \\
{\em $^5$ Department of Physics, Carnegie Mellon University, 5000 Forbes Ave, Pittsburgh,
PA 15232}\\
{\em $^6$ Department of Astronomy, University of Chicago, 5640 South Ellis
Ave, Chicago, IL 60637}\\ 
\vspace{0.6cm}
\end{center}
\normalsize
\begin {abstract}
We study the relations between stellar mass, star formation history,
size and internal structure for a complete sample of 122,808 galaxies 
drawn from the Sloan Digital Sky Survey. We show that low-redshift 
galaxies divide into two distinct families at a stellar mass of $ 3 
\times 10^{10} M_{\odot}$. Lower mass galaxies have young stellar
populations,   
low surface mass densities, and the low concentrations typical of      
disks. Their star formation histories are  more strongly correlated
with surface mass density than with stellar  mass.  A significant fraction of
the lowest mass galaxies in our sample have experienced recent starbursts. 
At given stellar mass, the sizes of low mass galaxies are log-normally distributed 
with dispersion $\sigma(\ln R_{50}) \sim 0.5$, in excellent agreement
with the idea that they form with little
angular momentum loss  through cooling and condensation in a 
gravitationally dominant dark matter halo. 
Their median stellar surface 
mass density scales with stellar mass as $\mu_* \propto M_*^{0.54}$, 
suggesting that the stellar mass of a disk galaxy is proportional to the
three halves power of its halo mass. 
All this suggests that the efficiency of the conversion of
baryons into stars in low mass galaxies  increases
in proportion to halo mass, perhaps as a result of supernova feedback processes.
At stellar masses above $3 \times 10^{10} M_{\odot}$, there is a 
rapidly increasing fraction of galaxies with old stellar populations, 
high surface mass densities and the high concentrations typical          
of bulges. In this regime, the size distribution remains log-normal, 
but its dispersion decreases rapidly with increasing  mass and
the median stellar mass surface density is approximately constant.                     
This suggests that the star formation efficiency decreases in the highest mass halos, and that 
little star formation occurs in massive galaxies after they have assembled.   

\end {abstract}
\vspace {0.8 cm}
\end {titlepage}
\normalsize

\section{Introduction}

Perhaps the most remarkable aspect of galaxies is their regularity -- the
fact that they can be classified  into well-defined sequences.
Hubble (1926) introduced the first scheme to categorize galaxies 
according to morphological type, and its basic concepts are still in
use today. In its simplest form, three basic galaxy types are
recognized: ellipticals, spirals and irregulars. These can be arranged
in a linear sequence along which many properties vary coherently. The 
most striking trends are the variations in morphology (featureless 
ellipsoid to spiral disk to irregular appearance) and in colour 
and star formation activity. These trends were so striking that they  
formed the basis of Hubble's original scheme. Elliptical galaxies have red
colours, little gas and no ongoing star formation; irregulars have blue
colours, large gas fractions  and are actively forming stars. An 
up-to-date review of star formation along the Hubble sequence is given 
in Kennicutt (1998).

It should be remembered that Hubble introduced his galaxy classification
scheme at a time when distance estimates were available for only a 
handful of galaxies. Hubble could not study the properties of 
galaxies as a function of parameters that measure their {\em absolute 
scale}: sizes, absolute magnitudes and masses. Today it is known that
the distribution of Hubble types depends strongly on galaxy luminosity. 
The typical elliptical galaxy is a factor $\sim 10$  more luminous
in the B-band than the typical irregular galaxy (Roberts \& Haynes 1994).
As a result, the faint end of the galaxy luminosity function 
is dominated by late-type galaxies with strong emission lines, 
the bright end by early-type galaxies with little star-formation (e.g.
Loveday et al 1992; Marzke et al 1994; Lin et al 1996; Zucca et al
1997; Marzke et al 1998; Christlein 2000). These trends in star
formation activity become even more pronounced when studied in the  
near-infrared.  This has led to suggestions that the ratio of present 
to past-averaged star formation rate may depend primarily on stellar 
mass and only secondarily on morphological type (Boselli et al 2001;
Balogh et al 2001).

There have been many studies of stellar populations and star formation    
histories that focus on galaxies of a particular Hubble type.
The star formation rates in most irregular and dwarf galaxies in the 
Local Group do not appear to have varied smoothly with time, but instead
have undergone large fluctuations or ``bursts'' (see Grebel 2000 for 
a recent review). There is evidence that the recent star formation 
histories of spirals depend on the surface brightness of their disks.
Low surface brightness galaxies appear to have bluer colours and  
higher gas mass fractions than high surface brightness galaxies 
(e.g. de Blok, van der Hulst \& Bothun 1995; de Blok, McGaugh \& van der 
Hulst 1996; Bell \& de Jong 2000). 
The situation for ellipticals is currently  unclear. Although there 
is considerable  evidence that the absorption line strengths of 
early-type galaxies vary systematically with their velocity
dispersions, there is much debate as to whether this reflects 
trends in  stellar age, in metallicity  or in the relative abundance of
different heavy elements (e.g. Trager et al 1998; Kuntschner et al 2001). 

The Sloan Digital Sky Survey will obtain $u,g,r,i$, and $z$-band 
photometry, spectra and redshifts for at least 700,000 galaxies down to 
a limiting $r$-band magnitude of 17.77. This is the most ambitious
survey of the local galaxy population ever undertaken and motivates    
a critical re-evaluation of trends in the star formation histories of        
galaxies as a function of mass, size, internal structure and
environment.

We have recently developed a new method to constrain the star
formation history, and to estimate the dust attenuation and stellar 
mass of galaxies (Kauffmann et al 2002; hereafter Paper I).
It is based on two stellar absorption line indices, the 4000 \AA \hspace{0.1cm}
break strength and the Balmer absorption line index H$\delta_A$.
These indices allow us to constrain the mean stellar age  
of a galaxy and the fraction of its stellar mass formed in bursts over 
the past few Gyr. A comparison with broad band magnitudes  then yields 
estimates of dust attenuation and of stellar mass.
We have generated a large library of Monte Carlo realizations 
of different star formation histories, including starbursts of 
varying strength. We have used this library to generate 
likelihood distributions for parameters such as
burst mass fraction, dust attenuation strength, stellar mass 
and stellar mass-to-light ratio for every galaxy in a complete
sample of 122,808 objects drawn from the Sloan Digital Sky Survey.

Paper I introduced our method and used it to measure how the total
stellar mass of the Universe is distributed over galaxies as a function 
of their stellar mass, size, concentration, colour and surface mass density.
In this paper, we focus on the {\em mass dependence} of the star
formation histories, sizes and structural parameters of galaxies.
We study to what extent the recent star formation histories of 
galaxies of different stellar mass can be characterized as continuous 
or as burst-dominated. We also show how the distributions of
structural parameters (size, surface mass density, concentration
index) vary with stellar mass, and we demonstrate that variations 
in star formation history and in structural parameters are tightly 
coupled. Finally, we interpret our findings in the context
of current theories of galaxy formation.

\section {Review of the Observed and Derived Parameters of Galaxies} 

The sample of galaxies analyzed in this paper is drawn from the Sloan 
Digital Sky Survey (York et al 2000; Stoughton et al 2002). We have 
included all galaxies with spectroscopic observations in the Data Release One (DR1)
and with Petrosian $r$ band magnitudes in the range 
$14.5 < r^* < 17.77$ after correction for foreground galactic extinction. 
Details about the spectroscopic target selection may be found in 
Strauss et al (2002). These data represent roughly 20\% of the 
projected survey total and include a total of 122,808 galaxies.

The reader is referred to Paper I for more information about the 
sample and a detailed description of the methods used to derive 
parameters such as stellar mass and burst mass fraction. We also discuss the
uncertainties in our methods in Paper I.  Here we 
provide a brief summary of the quantities that are studied in this paper:

\begin{enumerate}

\item {\bf Absolute magnitudes:}
All magnitudes quoted in this paper are Petrosian magnitudes.
In order to convert from apparent  to absolute magnitude, we assume 
a Friedman-Robertson-Walker cosmology with 
$\Omega=0.3$, $\Lambda=0.7$ and H$_0$= 70 km s $^{-1}$ Mpc$^{-1}$.
We have calculated the K-corrections $K(z)$ for each galaxy using the routines
in {\tt kcorrect v1\_11} (Blanton et al. 2002). In order to minimize the errors
in this procedure, we K-correct the magnitudes of all galaxies in our sample to $z=0.1$ 
More details about the SDSS photometric system may be found in 
Fukugita et al (1996) and Smith et al (2002). Information relating to the SDSS
camera and photometric monitoring system can be found in 
Gunn et al (1998) and Hogg et al (2001). A technical paper describing
the astrometry in the survey has recently been published by Pier et al (2002).

\item {\bf Stellar spectral indices:}
We have adopted the narrow definition of the 4000 \AA \hspace{0.1cm} break
introduced by Balogh et al (1999), which we denote D$_n$(4000).
We also use the Balmer absorption index H$\delta_A$ as defined by Worthey \&
Ottaviani (1997) using a central bandpass bracketed by two continuum 
bandpasses. The evolution of the two indices have been calibrated using a new
population synthesis code that incorporates high resolution stellar libraries
(Bruzual \& Charlot 2002). Our measurements of H$\delta_A$ are corrected for
contamination due to nebular emission.  As discussed in Paper I, the D$_n$(4000) index is an excellent  
age indicator for young ($< 1$ Gyr, D$_n$(4000)$< 1.5$) stellar populations, 
but for older stellar populations, the index depends quite strongly on 
metallicity. Strong Balmer absorption occurs in galaxies that  
experienced a burst of star formation 0.1-1 Gyr ago. 
Analyzed together, the two indices are a powerful probe of the recent star
formation  history of a galaxy. Because each is defined over a 
narrow wavelength interval, they are not sensitive to 
dust attenuation effects.

\item {\bf Stellar masses:} The stellar masses are derived assuming 
a universal initial mass function (IMF) in the parametrisation of 
Kroupa (2001). The typical 95\% confidence range in our estimated stellar masses
is $\pm$ 40 \%. 

\item {\bf Burst mass fractions:} We define the parameter $F_{burst}$ 
as the fraction of the total stellar mass of a galaxy that formed in 
``burst'' mode over the past 2 Gyr. In Paper I we showed that galaxies 
with continuous star formation histories occupy a very narrow locus 
in the H$\delta_A$/D$_n$(4000) plane. Galaxies that experienced recent 
bursts have H$\delta_A$ values that are displaced from this locus. 
Because the typical observational error on the H$\delta_A$ index is
large (1.4 \AA \hspace{0.1cm}) in our sample,  the median value of the likelihood
distribution of $F_{burst}$ is  sensitive to the mix of star
formation histories in our model library, which functions as a Bayesian prior for
our analysis. As a result, we focus below on galaxies for which our analysis implies
$F_{burst}>0$ with high confidence (97.5\% for our standard prior). The tests
described in Paper I showed the resulting sample to be insensitive to
the actual prior adopted. 

\item {\bf Galaxy sizes:} We study the distribution functions of R50($z$), the
radius enclosing 50\% of the Petrosian $z$-band luminosity  of a galaxy.
We note that the sizes output by the current SDSS photometric pipeline have 
not been corrected for seeing effects (Stoughton et al 2001).

\item {\bf Concentration indices:} In order to be consistent with previous
work (Blanton et al 2001; Strateva et al 2001; Shimasaku et al 2001) 
we define the concentration index $C$ as the
ratio $R90/R50$, where R90 and R50 are the radii enclosing 90\% and 50\% of the
Petrosian $r$-band luminosity  of the galaxy. 
It has been shown
by Shimasaku et al (2001) and Strateva et al (2001) that for bright galaxies, there
is a  good correspondence between concentration parameter and `by-eye' classification
into Hubble type,  with $C \sim 2.6$ marking the boundary between                                 
early and late type galaxies.  

\item {\bf Surface brightnesses and surface mass densities:}
Following Blanton et al (2001), we define the half-light surface 
brightness $\mu50$ to be the average $r$-band surface brightness 
within the $r$-band half-light radius R50 in magnitudes per square 
arcsecond. The main spectroscopic survey is complete down to a
limiting half-light surface brightness  $\mu50 =24.5$ mag 
arcsec$^{-2}$. We define the surface mass density $\mu_*$ as 
$0.5M_*/[\pi R50^2(z)]$, where $R50(z)$ is the Petrosian half-light 
radius in the $z$-band (we choose a radius defined in the $z$-band 
rather than $r$-band, as this provides a better approximation
to the radius enclosing half the total stellar mass).

\end{enumerate}

In this paper, we will often use bivariate density distributions $\phi(x,y)$ in
the parameters $x$ and $y$, defined such that $\phi(x,y) dx \hspace{0.1cm} dy$
is the number of galaxies per unit volume with $x$ in the interval
$(x, x+dx)$ and $y$ in the interval $(y,y+dy)$. Such bivariate 
distributions are calculated by weighting each galaxy in the sample
by $1/V_{max}$, where $V_{max}$ is the volume corresponding to the
total redshift range over which the galaxy would pass our sample 
selection criteria. Because our sample is extremely large, errors due to
large scale structure are expected to be small. In many cases we will be interested in trends in 
the distribution of the parameter $y$ as a function of the parameter
$x$ (for example in the distribution of concentration or size as a
function of stellar mass). In such cases, we make the trends more
visible by plotting the conditional distribution of $y$ given $x$, i.e.
$\phi(x,y)/\int\phi(x,y^\prime)dy^\prime$.

\section {The Observed Correlations} 
\subsection {The Dependence of Star Formation History on Stellar Mass}

In Fig. 1, we present the conditional density distributions of our 
two stellar age indicators D$_n$(4000) and H$\delta_A$ as a function 
of stellar mass and as a function of $g$-band absolute magnitude.  
The grey-scale indicates the fraction of galaxies in a given 
logarithmic mass (or magnitude) bin that fall into each age-indicator
bin. The contours are separated by factors of 2 in population density.

It is clear that D$_n$(4000) and H$\delta_A$ depend strongly on both
stellar mass and  absolute magnitude. The main effect of
transforming from luminosity to mass is to produce a more regular
variation of the indices. Fig. 1 shows that low mass galaxies are `young'-- they
typically have low values of D$_n$(4000) and high values of H$\delta_A$. 
At a mass of $M_* \sim 3 \times 10^{10}$ M$_{\odot}$, 
a transition towards older stellar populations begins to take effect. 
Almost all of the most massive galaxies in our sample have  high values of D$_n$(4000)
and low values of H$\delta_A$.

Fig. 2 shows ``slices'' through the D$_n$(4000)-$M_*$ distribution.
Recall that the measurement error on the D$_n$(4000) index is 
small, typically around 0.04, or a few percent of the total 
range of values spanned by galaxies in our sample. Again, one sees 
a striking trend towards older stellar populations for galaxies with 
larger stellar masses. Both in Fig. 1 and in these plots one gets the impression
of two separate and relatively well-defined populations. The relative
weight of the `old' population increases strongly at stellar masses
above $\sim 10^{10} M_{\odot}$.

The bivariate density distribution of H$\delta_A$ and D$_n$(4000) 
is shown in Fig. 3 for galaxies in 8 ranges  of stellar mass spanning     
four orders of magnitude from $M_* = 10^{8} M_{\odot}$ to $M_* = 
10^{12} M_{\odot}$. In these plots, the grey scale indicates the
fraction of galaxies that fall into each bin of $H\delta_A$ and D$_n(4000)$.
It is  striking how galaxies move diagonally across the H$\delta_A$/D$_n$(4000)
plane as their masses increase.

From Fig. 3, one can see clearly that low mass galaxies
with $D_n(4000) < 1.5$ (characteristic of stellar populations with mean ages
of less than a few Gyr) have a larger scatter in H$\delta_A$ equivalent width 
than high mass galaxies with similar break strengths. This means that 
the fraction of low mass galaxies that have experienced recent bursts 
is higher than that for high mass galaxies,                                    
even when the two populations are compared at similar mean stellar age.
We illustrate this in detail in  Fig.4, where we plot the median value
of H$\delta_A$ as a function of D$_n$(4000) for low mass and for high
mass galaxies. The solid and   
dotted errorbars indicate the 25th to 75th and 5th to 95th percentile ranges of
the distribution of H$\delta_A$ values. 
For comparison, we have
overplotted 200 different model galaxies with continuous
star formation histories. The models span a  wide  range in  
exponential timescale ($\tau=0.1$ to 10 Gyr)  and formation redshift ($z_{form}=20$ to 0.2).
The reader is referred to Paper I for more details about the models.
Solar metallicity models are plotted in the right hand panel. In the left hand panel, we have
plotted 20\% solar models, which are more appropriate for galaxies with masses
between $10^8$ and $10^9$ $M_{\odot}$ (Tremonti et al 2003)
For massive galaxies , the data                                             
agree extremely well with the model predictions. The scatter in H$\delta_A$ at given D$_n$(4000)
is consistent with that expected from observational errors.
Note that although we plot the distribution of H$\delta_A$ down to very low D$_n$(4000) values,
 Fig. 2 shows that less than 5\% of galaxies with stellar masses in the range
$10^{10}-10^{11} M_{\odot}$ have values D$_n$(4000)  too low to be consistent
with any of our continuous models.
For low mass galaxies, this is no longer true. Fig. 2 shows that  $\sim 30$\% of all galaxies
have D$_n$(4000)$<1.2$. As discussed in Paper I, many of these galaxies are likely to
be experiencing a starburst at the present day.  Fig. 4 shows that for low mass galaxies,
the data are also displaced upwards  relative to the predictions of
the continuous models and the scatter is too large to be consistent with
observational errors. As discussed in Paper I, galaxies with moderate values of D$_n$(4000)
and very high values of H$\delta_A$ are likely to be in a post-starburst phase.

In Paper I, we introduced models in which galaxies formed stars 
in two different modes: i) a ``continuous'' mode , which we parametrized
by an exponential law with timescale $\tau$ and starting time $t_{form}$, and
ii) a superposed ``burst'' mode.              
A burst was defined to be an episode of star formation lasting
between $3\times 10^7$ and $3\times 10^8$ years in which a fraction
$F_{burst}$ of the total stellar mass of the galaxy was formed.
In Paper I, we also introduced a method that allowed us to calculate the
{\it a posteriori} likelihood distribution of $F_{burst}$                     
for each galaxy in our sample, given its observed absorption line
indices and the measurement errors on these indices.
In Fig. 5, we plot the distribution of galaxies as a function of 
$F_{burst}$ for  8 different mass ranges.
The distribution of the median value of $F_{burst}$ (which we denote $F_{burst}(50\%)$) for all galaxies
in the given mass range is plotted as a solid histogram.
We define a subsample of high-confidence bursty galaxies as those objects with $F_{burst}(2.5\%)>0$, where
$F_{burst}(2.5\%)$ is the lower 2.5 percentile point of the likelihood distribution of $F_{burst}$.
The distribution of $F_{burst}(50\%)$ for this subsample is plotted as a dotted histogram in Fig. 5.

In our lowest mass range ($10^8-3\times 
10^8 M_{\odot}$), we find that half of all galaxies have $F_{burst}(50\%) > 0$. However, 
it is only for 10\% of these objects that we can state with high 
($>$97.5\%) confidence that a burst {\em did occur} in the recent past.
The fraction of galaxies that have experienced recent bursts decreases
very strongly with increasing stellar mass. This trend is apparent both in
the fraction of galaxies with  $F_{burst}(50\%)>0$ and in the fraction
with $F_{burst}(2.5\%)>0$. There are very few high
confidence bursty galaxies in our high mass bins.  Out of our entire sample of 120,000
galaxies, we only pick out around 500 galaxies with masses comparable to that
of the Milky Way and with $F_{burst}(2.5\%) > 0$.
For our full sample, the median value of the  burst mass fraction
correlates with stellar mass. The same is not true for the high confidence bursty
galaxies -- these systems have burst mass fractions that are independent of
stellar mass. Note that our requirement that a burst be detected at high confidence
will, by definition, bias this subsample towards high burst fractions.
The trend in burst mass fraction for our full sample is in some sense expected, because
it is well known that more massive galaxies have
lower gas mass fractions and thus contain less fuel for star formation 
(e.g. Roberts \& Haynes 1994; Boselli et al 2001; Bell \& de Jong 2001).
What about the influence of our adopted prior? The analysis in Paper I showed that decreasing    
the number of bursty galaxies in the model library resulted in {\em lower} estimated values
of $F_{burst}(50\%)$. Our standard prior assumes that the fraction of bursty galaxies
is constant, in apparent contradiction with the results shown in Fig. 5.
Our choice of prior would
therefore tend to {\em weaken}, rather than strengthen, any true decrease in $F_{burst}$ 
towards high masses. Paper I also demonstrated that the definition and inferred properties
of our high confidence subset of bursty galaxies in insensitive to the adopted prior.

In Figs. 6 and 7 we compare the distributions of H$\alpha$  
emission line equivalent widths for galaxies with $F_{burst} (2.5\%)>0$ and   
for galaxies with $F_{burst}(50\%)=0$. In Fig. 6,  we have subdivided 
the two samples into different bins in D$_n$(4000) in order to take 
into account any systematic mean age difference between galaxies with 
and without recent bursts.  For galaxies with low values of D$_n$(4000), there is a tendency for
bursty galaxies to have {\em
stronger} H$\alpha$ emission than non-bursty  galaxies with  similar break strengths.
These are galaxies whose present-day  star formation rates are enhanced over other systems of
similar mean stellar age (recall that the H$\alpha$ emission is coming from stars less
than $10^7$ years in age).
For galaxies with high values of D$_n$(4000), the trend is reversed.
These high D$_n$(4000) galaxies are presumably ``post-starburst'' systems that experienced
a strong burst some time in the past, but have subsequently stopped forming any stars.
In Fig. 7, we compare galaxies with $F_{burst}(2.5\%) >0$ and galaxies with
$F_{burst}(50\%)=0$ in fixed bins of stellar mass. Low mass bursty  galaxies
are skewed to higher H$\alpha$ equivalent widths, while high mass bursty galaxies
exhibit a wider range in equivalent widths than their counterparts with
$F_{burst}(50\%)=0$. Because information about  H$\alpha$ was not used when
selecting our sample of bursty galaxies, we regard these differences in emission line properties
as evidence that our method does pick out
a sample of galaxies with very strong variations in recent star formation
history.

\subsection {The Dependence of Structural Parameters on Stellar Mass}

In Fig. 8, we present conditional density distributions for concentration 
index $C$, $r$-band half-light surface brightness $\mu50$,
and surface mass density $\mu_*$ as functions of stellar mass 
and of $r$-band absolute magnitude. Fig. 8 demonstrates that the 
structural parameters of galaxies correlate both with absolute
magnitude and with stellar mass, but once again we find that the 
trends are smoother when plotted as a function of stellar mass.

The surface mass density exhibits a strikingly tight correlation 
with stellar mass. $\mu_*$ increases by nearly two orders of 
magnitude from $\sim 10^7 M_{\odot}$ kpc$^{-2}$ for galaxies with 
$M_* \sim 10^8 M_{\odot}$ to $\sim 10^9 M_{\odot}$ kpc$^{-2}$ for
the most massive galaxies with $M_*=10^{12} M_{\odot}$. In contrast, 
the $r$-band surface brightness only increases by a factor of $4$ 
as $M(r)$ increases by 8 magnitudes. There is a sharp change in slope in the 
$\mu_*-M_*$ relation at a stellar mass of $\sim 3\times 10^{10} 
M_{\odot}$. A  transition at the same stellar mass is also seen
in the $C-M_*$ relation. Interestingly, this is also  the  stellar 
mass at which galaxies switch from low D$_n(4000)$ to high D$_n$(4000) in Fig. 1.
It is important to note that virtually all galaxies in our sample 
with $M_* > 10^8 M_{\odot}$ have $\mu_{1/2} < 23.5$ mag
arcsec$^{-2}$. The main  spectroscopic survey is complete to surface 
brightnesses a magnitude fainter than this, so our results should not
be biased by surface brightness selection effects.    

Fig. 9 presents the size distributions for galaxies in eight disjoint ranges 
of stellar mass. We plot the fraction of galaxies as a function of 
the natural logarithm of the half-light radius in the $z$-band. 
These size distributions are extremely well described  by a log-normal 
function
\begin {equation}  P(R)dR = \frac {1}{\sqrt{2 \pi} \sigma} \exp
\left[ - \frac {(\ln (R/R_{med}))^2}{2 \sigma^2} \right]   
\frac{dR}{R}, \end {equation}
where $R_{med}$ is the median value of the size distribution and
$\sigma$ is the dispersion in $\ln R$. We have indicated the value 
of $\sigma$ and $R_{med}$ for the lognormal fit in each panel in 
Fig. 8. The {\em shape} of the size distribution is largely independent 
of mass for galaxies with $M_* <  3 \times 10^{10} M_{\odot}$ and 
is well fit by $\sigma \sim 0.4-0.5$. The median size $R_{med}$
scales with stellar mass as  $R_{med} \propto M_*^{0.18}$. For stellar masses
larger than $ 3 \times 10^{10} M_{\odot}$, the shape of the
size distribution is still lognormal, but the dispersion decreases. 
$R_{med}$ also increases more rapidly with mass ($\propto M_*^{0.33}$)
in this regime.  

Figs. 10 and 11 show the distributions of concentration index $C$ and 
surface mass density $\mu_*$ for these same eight stellar mass ranges.
The distribution of $C$ is nearly independent of stellar mass  for
galaxies with  $ M_* < 3 \times 10^{10} M_{\odot}$. The median value of $C$
for these systems is around 2.3. Only about  10\% of low mass
galaxies have $C>2.6$, the value that marks the transition from 
late-type to early-type morphologies (Strateva et al 2001).
At larger masses, the distribution
shifts to progressively higher concentrations. In our highest mass bin 
90\% of galaxies have $C$ index values larger than 2.6. 
Fig. 11 shows that for galaxies with $M_* < 3 \times 10^{10} M_{\odot}$,
the median surface mass density scales with stellar mass
as $\mu_* \propto M_*^{0.63}$. The $\mu_*$ distribution is fairly broad 
and its shape does not appear to depend very strongly on stellar mass.
At values of $M_*$ larger than $3 \times 10^{10} M_{\odot}$, the scaling    
of the median value of $\mu_*$ with stellar mass becomes much weaker. 
The shape of the distribution
function skews  and is eventually strongly peaked at values of $\mu_*$ 
around $10^9 M_{\odot}$ kpc$^{-2}$.

We note that all structural parameters discussed in this section
have been calculated within circular apertures.
We have estimated the additional scatter in size resulting from variations in
axis-ratio and find that this has very small ($< 10\%$) effect
on the values of the dispersion $\sigma$ that we estimate.
Because of seeing effects, $R50(z)$ may be overestimated for galaxies with small
angular sizes.  We have split the galaxies in each stellar mass range into two 
equal samples according to redshift, and we have compared the size distribution of
the nearby sample with that of the full sample. This is illustrated in two
of the panels of Fig. 9. As can be seen, the effects appear to be quite small. 
In addition, we have checked the scalings using  $R90(z)$ and find that our 
conclusions remain unchanged.

\subsection{ The Connection Between Star Formation History and 
Structural Parameters}

In the previous two subsections we showed that both the star formation 
histories and the structural parameters of galaxies depend strongly on 
stellar mass. It is therefore not surprising that the star formation 
histories of galaxies correlate with $\mu_*$ and $C$. This is
illustrated in Fig. 12 where we present the conditional density 
distributions of the two stellar absorption indices D$_n$(4000) 
and H$\delta_A$ as functions of surface mass density and of
concentration. Galaxies with low surface density have low values of D$_n$(4000)
and galaxies with high surface density have 
high values of D$_n$(4000), with a strong transition at $\mu_* \sim 
3\times 10^8 M_{\odot}$ kpc$^{-2}$. There is also a striking transition
in the values of D$_n$(4000) and H$\delta_A$ at $C$ values around 2.6.
We note that this corresponds very well to the  value of $C$ recommended 
by Strateva et al (2001) for optimum separation between early and 
late Hubble types, based on an analysis of a small sample of 
galaxies classified by eye.

We now ask whether it is possible to pinpoint the  {\em primary} 
cause of variations in the star formation history of galaxies -- 
do galaxies have low values of D$_n$(4000) and high values of
H$\delta_A$ because their masses are 
small or because their surface mass densities are low? We attempt to 
address this question in Fig. 13  where we plot the
fractions of galaxies with  D$_n$(4000)$> 1.55$ and with H$\delta_A < 2.0$    as a function 
of surface mass density in narrow ranges   of  stellar mass (left panels), and as 
a function of stellar mass in narrow ranges of surface mass density (right panels). 
We only plot these fractions for values of  $M_*$ and $\mu_*$ where there are
more than 100 galaxies per bin. 
D$_n$(4000)=1.55 and H$\delta_A$=2 are  chosen 
as natural places to divide the sample, because this is where 
clear transitions occur in the D$_n$(4000)/$C$ and H$\delta_A$/$C$  plots in Fig. 12.  
(Note that the $C$ index is determined directly from the photometric data, independently
of both stellar mass and surface mass density).      

Fig. 13  shows that the two indices are more fundamentally related to
$\mu_*$ than to $M_*$. The fraction of galaxies with D$_n$(4000)$>1.55$ or with
H$\delta_A<2$ is  largely
independent of stellar mass at a given value of $\mu_*$, although the most massive galaxies
with $M_* > 10^{11} M_{\odot}$ clearly deviate 
in that they have                                  
larger values of D$_n$(4000) and smaller values of 
H$\delta_A$ at given $\mu_*$. At given stellar mass, however, the fraction
of galaxies with D$_n$(4000)$>$1.55 or with H$\delta_A < 2$  is a strongly increasing function of
$\mu_*$.
These results are in qualitative agreement with those of Bell \& de Jong (2000),
who studied trends in the  optical and infrared colours 
of a sample of nearby {\em spiral} galaxies and found
that the  colours  most sensitive to star formation history
 correlate best with surface density.
It also fits in well with numerous studies that have demonstrated a clear correlation
between star formation {\em rate} and gas surface density for samples
of spirals (e.g. Kennicutt 1983;
Wong \& Blitz 2002).

\subsection {Summary}

In the previous subsections, we demonstrated that there are strong correlations
between the star formation histories, stellar masses and structural parameters
of galaxies. Here we provide a summary of our results:

\begin {enumerate}
\item Low mass galaxies have low values of D$_n$(4000) and high values
of H$\delta_A$, indicative of young stellar populations . High mass
galaxies have  high values of D$_n$(4000) and low values of
H$\delta_A$, indicative  of old stellar populations. A  sharp transition from young 
to old occurs at a stellar mass of around $3 \times 10^{10} M_{\odot}$.
\item Low mass galaxies have low surface mass densities and low 
concentrations. High mass galaxies have high surface mass densities 
and high concentrations. An abrupt change in the slopes of the 
$\mu_*-M_*$ and $C-M_*$ relations occurs at a stellar mass of 
$ 3 \times 10^{10} M_{\odot}$.
\item The size distribution function of galaxies less massive than 
$ 3 \times 10^{10} M_{\odot}$ is well characterized by a lognormal 
distribution with dispersion $\sigma=0.4-0.5$. Galaxies more massive
than $3 \times 10^{10} M_{\odot}$ also have a lognormal size 
distribution, but the dispersion is smaller.                           
\item Below the transition mass the median size of galaxies increases
as $M_*^{0.18}$. Above the transition mass the increase is more rapid,
$R\propto M_*^{0.3}$.
\item  Galaxies with low surface mass densities and low concentrations have
D$_n$(4000) and H$\delta_A$ values indicative of 
young stellar populations.  Galaxies with high surface mass densities and
high concentrations have D$_n$(4000) and H$\delta_A$ values indicative of
old stellar populations. The transition from
young to old stellar populations occurs at $\mu_* \sim 3\times 
10^8 M_{\odot}$ kpc$^{-2}$ and $C \sim 2.6$.
\item The star formation histories of low mass  galaxies, as traced by 
D$_n$(4000) and H$\delta_A$ ,  appear to be more fundamentally
related to surface mass density than to stellar mass.
\item  The fraction of galaxies that have experienced recent
starbursts decreases strongly with increasing stellar mass (and increasing surface mass density). 

\end {enumerate}

\section {Interpretation and Discussion} 

Our analysis has demonstrated that there is a sharp transition in 
the physical properties of galaxies at a stellar mass of 
$ \sim 3\times 10^{10} M_{\odot}$. Galaxies less massive
than this have low surface mass densities, low concentration 
indices typical of disks  and young stellar populations. More massive galaxies
galaxies have high surface mass densities, high concentration indices 
typical of bulges, and predominantly old stellar populations.

The size distribution of disk galaxies is of particular interest, because 
it can be compared directly to simple theoretical models. In a 
classic paper in 1980, Fall \& Efstathiou considered the formation 
of a disk by the condensation of gas in a gravitationally
dominant dark matter halo. If the gas initially has the same specific 
angular momentum as the dark matter and conserves its angular 
momentum during its contraction, then the characteristic radius
of the disk will scale as
\begin {equation} R_d \propto \lambda R_{halo}, \end {equation} 
where $R_{halo}$ is the virial radius of the halo. The spin
parameter $\lambda$ is defined as
\begin {equation} \lambda = J |E|^{1/2}  G^{-1} M_{halo}^{-5/2}, \end {equation}  
where $E$ is the total energy of the halo, $J$ is its total angular
momentum and $M$ is its total mass (see Mo, Mao \& White (1998) for a
detailed analytic model).  

N-body simulations show that in hierarchical clustering cosmologies,  
the distribution of $\lambda$ for dark matter halos has lognormal form
with parameters which depend very weakly on cosmology or on halo mass 
(Barnes \& Efstathiou 1987; Warren et al 1992; Cole \& Lacey 1996;
Lemson \& Kauffmann 1999) with typical values $\lambda_{med} 
\simeq 0.04$ and $\sigma_{\lambda} \simeq 0.5$. The fact that 
for low mass galaxies the shape and width of the size distributions 
in Fig. 9 agree so well with this simple theory is truly remarkable.
 We note that our results are in qualitative agreement with
those of Syer, Mao \& Mo (1999), who find that the distribution of spin parameter
$\lambda$ inferred from a sample of nearby disk galaxies is in excellent
agreement with the predictions of cosmological simulations. On the other hand,
our results appear to disagree with those of de Jong \& Lacey (2000), who
find that the distribution of disk sizes is narrower than predicted by
simple models. We note that neither of these two studies  addressed possible
variations in the size distributions of disk galaxies as a function of
luminosity or of mass.

It is not obvious how the stellar mass of a galaxy should scale
with the mass of its dark matter halo. For simplicity, one might
assume that a fixed fraction of the baryonic mass ends up in disk stars, 
in which case $M_* \propto R_{halo}^{3}$ and  $R_d \propto M_*^{1/3}$ 
(this assumes that baryons follow the dark matter, at least initially).
From Fig. 9, we find that median size  increases {\em more slowly} than predicted. 
This means that 
the stellar mass fraction must {\em increase} with halo mass. 

We now derive the scaling between the stellar mass of a disk galaxy
and the mass
of its dark matter halo from the {\em slope} of the $\mu_*-M_*$ relation.
If we adopt the  assumptions of the simple model, we can write \begin {equation} 
\mu_* \propto M_*/R_d^2 \propto \epsilon M_{halo}/R_{halo}^2 \propto 
\epsilon M_{halo}^{1/3}
\propto \epsilon^{2/3} M_* ^{1/3}, \end {equation}
where $\epsilon \propto  M_*/M_{halo}$ is the efficiency with which the disk has converted
the available gas into stars.
The Fall \& Efstathiou arguments are only directly relevant to the 
formation of disk galaxies, while the conditional density
distributions shown in                           
Fig. 9 are derived using all galaxies in the sample. At stellar masses
greater than  $10^{10}M_{\odot}$, a large fraction of the galaxies are
bulge-dominated systems. In order to clarify the scaling properties
of our two principal classes of galaxy, we define two subsamples which
eliminate most of the mixed systems: i) a sample of late-type galaxies 
with   $C<2.6$ and $D_n(4000) < 1.55$ (41312 galaxies); and ii) a sample 
of early-type galaxies with $C>2.6$ and $D_n(4000)>1.55$ (53119 galaxies). 
Fig. 12 demonstrates that these cuts divide young, disk-dominated
systems from old, bulge-dominated systems surprisingly cleanly.  

The $\mu_*-M_*$ relation of our late-type subsample is shown in the 
left panel of Fig. 14, while the right panel shows the 
corresponding relation for early-types.
It is striking that the $\mu_*-M_*$ relation for the late-type subsample can be
described quite well by a single power-law. The $\mu_*-M_*$ relation for 
early-type galaxies still appears to undergo a transition in slope.
(Note that there are few late-type galaxies with
stellar masses above $2\times 10^{11} M_{\odot}$ and few early-types with
$M_*$ below $10^{10} M_{\odot}$.)  Fig. 14 also shows that early-type 
galaxies of given stellar mass have higher stellar surface
densities than late-type galaxies of the same mass. This difference
is largest for the lower mass objects and disappears for the most
massive systems in our samples.

Fitting to the dominant part of the population of late-type galaxies (i.e. over the
stellar mass range $10^8$ --  $10^{11} M_{\odot}$) , we find $\mu_* \propto 
M_*^{0.535 \pm 0.03}$,  so that
$\epsilon \propto M_*^{0.30 \pm 0.05} \propto M_{halo}^{0.43 \pm 0.09}$ and $R_d\propto 
M_*^{0.23 \pm 0.02}$. The star formation efficiency for disk galaxies apparently
increases with halo mass.
Such an increase 
is expected in a model where supernova feedback controls the dynamics
of the interstellar medium. 
(e.g. McKee \& Ostriker 1977; Efstathiou 2000). In most feedback models,
supernovae inhibit the formation of stars
in low mass halos, leading to small values of $M_*/M_{halo}$ (e.g. White \& Frenk
1991; Kauffmann, White \& Guiderdoni 1993; Cole et al 1994).
However, because we do not have any information about the gas fractions of
the galaxies in our sample, we cannot say  whether the suppression of
star formation occurs because gas is expelled from the galaxy and from
its halo, as is often
assumed in these models,
or because the conversion of gas into stars
is slowed in low mass systems.
It is interesting that Tremonti et al (2003) find the gas phase metallicity of low mass galaxies
to increase with stellar mass as $Z \propto M_*^{0.4-0.5}$ (depending on which metallicity
indicator they use), suggesting a similar dependence of overall star formation
efficiency on stellar mass.

For early-type galaxies, we fit over the range $3 \times 10^{10}-3 \times 10^{12} M_{\odot}$
and obtain  $\mu_* \propto  M_*^{0.0 \pm 0.02}$. If we assume that $R_d \propto R_{halo}$
for these galaxies, this would imply that
$\epsilon \propto M_*^{-1/3}$, i.e. the
the star formation efficiency for ellipticals {\em decreases} with halo
mass. Clues as to why this is  the case may come from recent {\em Chandra} and
{\em XMM} observations, which show that cooling of  gas  to temperatures below  1 keV  
is suppressed at the centres of rich clusters, where the most
massive ellipticals are located (e.g. Fabian et al 2001; Bohringer et al 2001).
It seems very likely that the assembly of bulge-dominated galaxies 
involved chaotic processes such as mergers or inhomogenous, 
rapidly star-forming collapse, which, for most systems, must have
completed well before the present day. Under these circumstances, it is not
clear that our assumed scaling between  galaxy radius and 
dark matter halo radius is valid. Even if the simple scaling laws
were to apply, the halo sizes and surface
densities relevant for early-type galaxies are likely to be
defined at higher effective redshift than those relevant for
late-type galaxies. This may explain
why early-type galaxies in Figure 14 have higher surface densities 
than late-type galaxies of similar mass. The scatter in surface 
density (and size) is also smaller for early-type than for late-type galaxies,
as is very evident in Figure 14 as well as in Figures 8 and 9. 

We note that the Lyman Break Galaxies (LBGs) at high redshift can plausibly be identified with
the progenitors of early-type galaxies.
Pettini et al.(2001) have derived a mean effective dynamical mass 
of 1.3 $\times 10^{10} M_{\odot}$
within a mean half-light radius of 2.3 kpc for a sample of LBGs at $z \sim 3$.
 Shapley et al.(2001) and Papovich et al. (2001) have estimated the stellar massses
of LBGs using their observed 
rest-frame ultraviolet and optical spectral energy
distributions. These agree well with the dynamical mass estimates. 
Lyman Break Galaxies thus have characteristic
surface mass densities of $\sim 10^9 M_{\odot}$ kpc$^{-2}$,
quite similar to those of present-day early-type galaxies. Moreover, if the 
star-formation associated with the Lyman Break Galaxies is integrated from
from $z \sim$ 2 to 6, one finds that 
they could have produced most of the stars in                        
present-day early-type galaxies (Steidel et al. 1999; Thompson et al 2001). 

In the framework of hierarchical models of galaxy formation, the shape of the size distributions 
of low mass galaxies is easy to understand, but
the trend 
of younger mean stellar ages with decreasing stellar mass  is  
puzzling, since lower mass dark matter halos typically form
at earlier epochs. If all the gas cooled off and formed stars when these 
halos were assembled, lower mass galaxies would  have older        
stellar populations (e.g. Kauffmann, White \& Guiderdoni 1993). We 
have seen,however, that non-gravitational processes such as supernova feedback
may play an important role in regulating the rate at which stars form 
in these systems. To explain the observations, these processes must
slow the conversion of gas into stars so that low mass systems remain
gas rich until the present day and have low overall efficiencies for
turning gas into stars. We have seen that the star formation histories
of low mass galaxies are correlated more strongly with  surface density
than with stellar mass. If feedback processes are able to prevent the gas in  low
suface density galaxies  from dissipating to form dense molecular clouds, then their star formation
timescales will be long.

Conversely, it may appear puzzling that star formation has terminated in
the majority of high mass galaxies, because the hierarchical model claims that
these systems were
assembled relatively late. In order to explain the observations, 
star formation must have been  both efficient and rapid in the
progenitors of these systems, so that the final stages of their
assembly involved rather little dissipation or star formation. 
In current models, galaxies that fall into groups and clusters are 
stripped of their reservoir of gas. As a result, their star formation 
rates decline and their colours redden (see for example Diaferio et al 
2001). This is unlikely to explain why the {\em
majority} of massive galaxies do not form stars at the present day, 
because many  do not reside in rich groups and clusters.
In such systems, gas must somehow be prevented from cooling and
forming stars after the bulge has formed.
Again, the suppressed cooling at the centres of observed galaxy clusters may offer a clue to
the mechanism responsible for this.

Finally, it will be important to understand the physical origin of the
characteristic mass scale that is so strikingly imprinted
on the galaxy population. The structural properties and formation paths          
of dark matter halos vary smoothly with mass. The abrupt transition in the stellar ages,
star formation histories and structure of galaxies
that we find at $3 \times 10^{10} M_{\odot}$ can only be explained
by star formation and/or  feedback processes. 
We believe that this tension between simplified 
phenomenological models of galaxy evolution and the trends seen in the data 
will lead to new insight
into the physical processes that regulate how galaxies form and evolve.
We intend to address these issues in more detail in future work.

\vspace {1.5cm}

We thank David Weinberg for  helpful discussions and comments.
We also thank the anonymous referee for comments that helped improve
the paper.
S.C. thanks the Alexander von Humboldt Foundation, the Federal Ministry of Education
and Research, and the Programme for Investment in the Future (ZIP) of the German
Government for their support.

 The Sloan Digital Sky Survey (SDSS) is a
joint project of The University of
           Chicago, Fermilab, the Institute for
                              Advanced Study, the Japan Participation
                              Group, The Johns Hopkins University, Los Alamos
                              National Laboratory,  the
                              Max-Planck-Institute for Astronomy
                              (MPIA), the Max-Planck-Institute for
                              Astrophysics (MPA), New Mexico State
                              University, Princeton University, the
                              United States Naval Observatory, and the
                              University of Washington. Apache Point
                              Observatory, site of the SDSS
                              telescopes, is operated by the
                              Astrophysical Research Consortium (ARC).

                              Funding for the project has been
                              provided by the Alfred P. Sloan
                              Foundation, the SDSS member
                              institutions, the National Aeronautics
                              and Space Administration, the National
                              Science Foundation, the U.S. Department
                              of Energy, the Japanese Monbukagakusho,
                              and the Max Planck Society. The SDSS Web
                              site is http://www.sdss.org/.

\begin{figure}
\centerline{
\epsfxsize=16cm \epsfysize=14cm \epsfbox{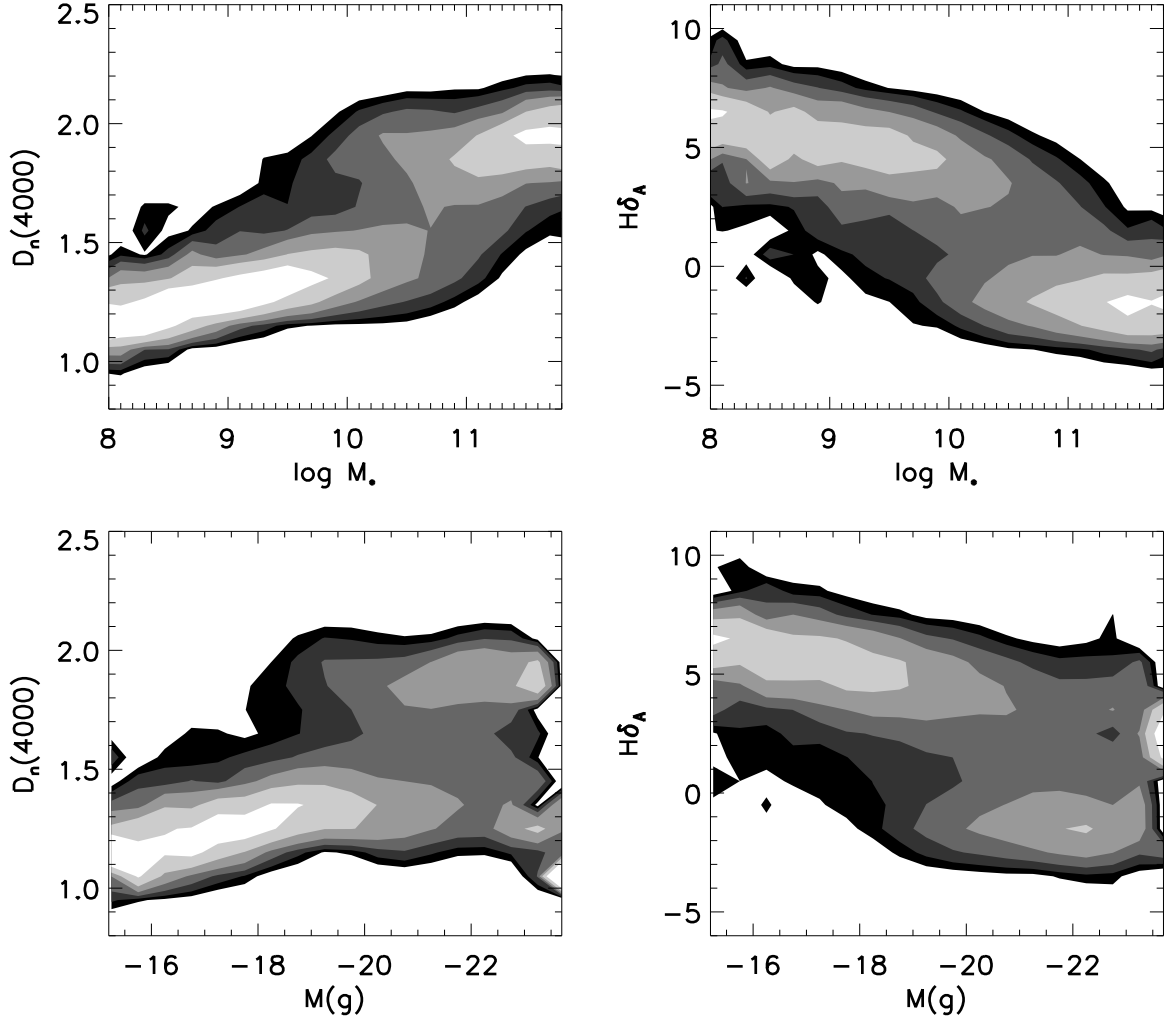}
}
\caption{\label{fig1}
\small
Conditional density distributions showing trends in the stellar age indicators
D$_n$(4000) and H$\delta_A$ as  functions of the logarithm of stellar
mass and of $g$-band absolute magnitude. Galaxies have 
been weighted by $1/V_{max}$ and the bivariate distribution function has
been normalized to a fixed number of galaxies in each bin of $\log
M_*$ or M($g$). Here and in all subsequent contour plots, each
contour represents a factor two change in density.}
\end {figure}
\normalsize

\begin{figure}
\centerline{
\epsfxsize=16cm \epsfbox{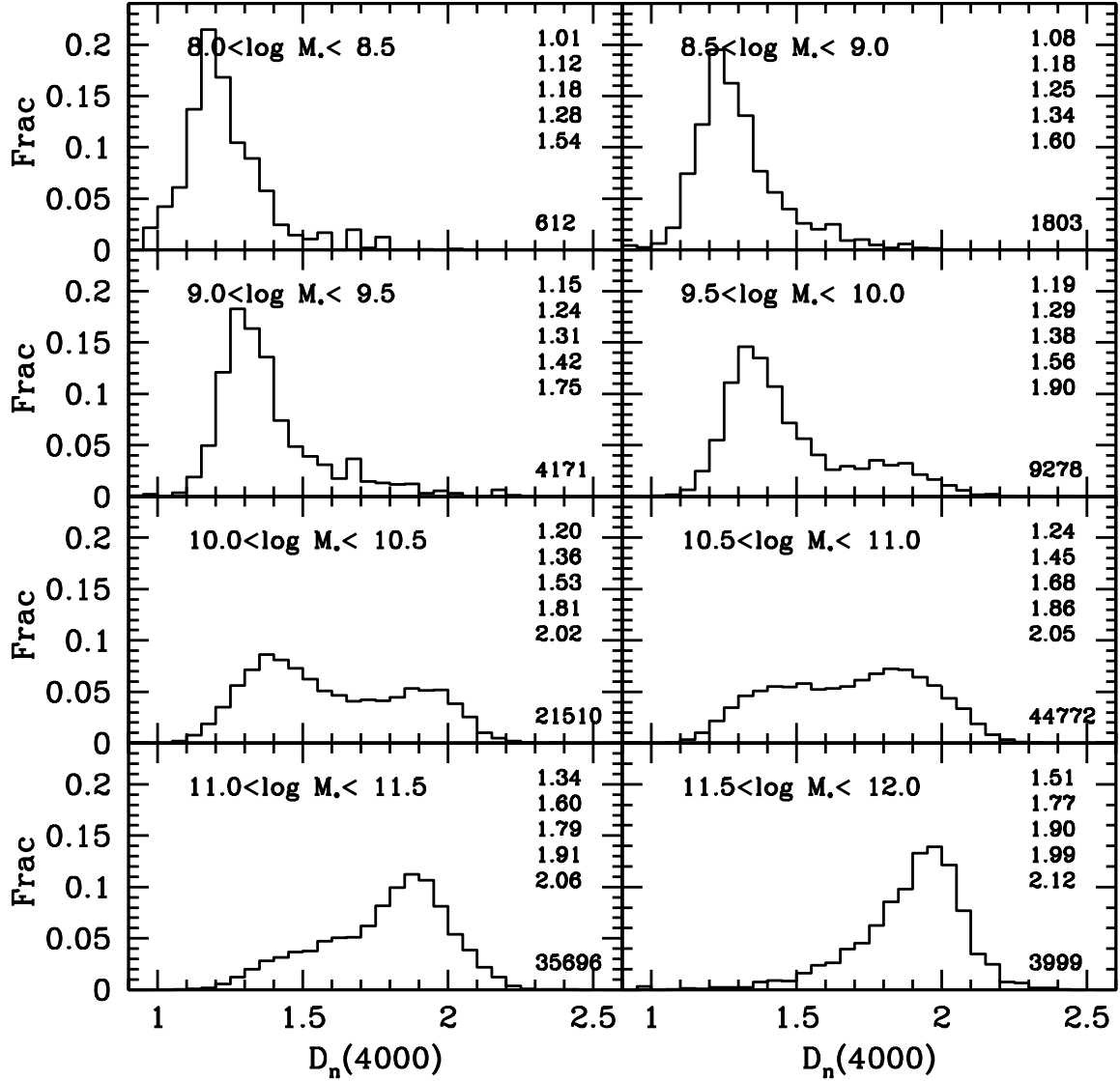}
}
\caption{\label{fig2}
\small
Histograms showing  the fraction of galaxies as a function of 
D$_n$(4000) in 8 different ranges of stellar mass. The numbers 
in the upper right corner of  each panel list, from top to bottom, the
5th, 25th, 50th, 75th and 95th percentiles of the distribution. The number
in the lower right corner is the number of galaxies contributing to
the histogram.}
\end {figure}
\normalsize

\begin{figure}
\centerline{
\epsfxsize=14.5cm \epsfysize=18cm \epsfbox{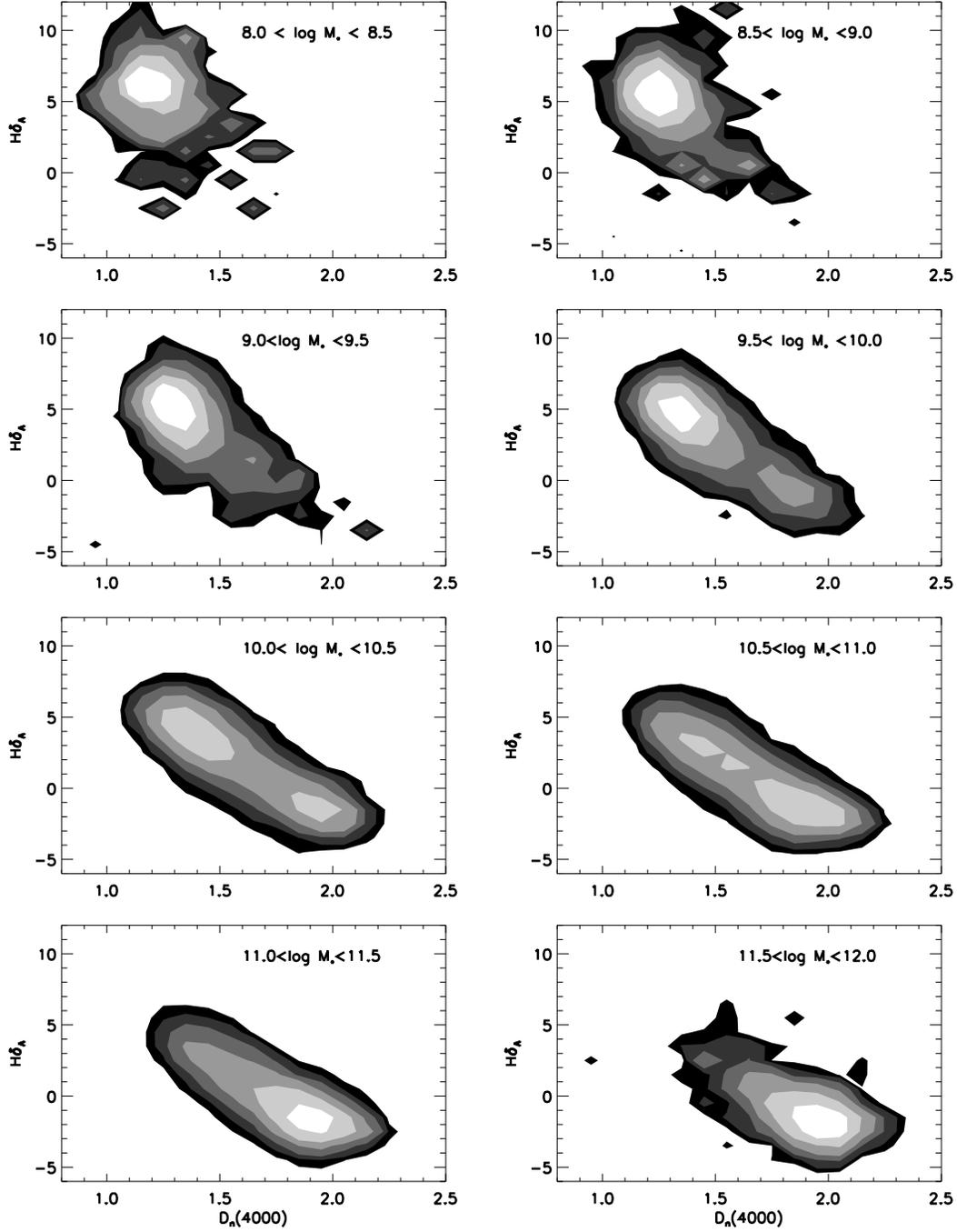}
}
\caption{\label{fig3}
\small
The bivariate distribution function of H$\delta_A$ and D$_n$(4000) 
in 8 ranges of stellar mass.}
\end {figure}
\normalsize

\begin{figure}
\centerline{
\epsfxsize=16cm \epsfbox{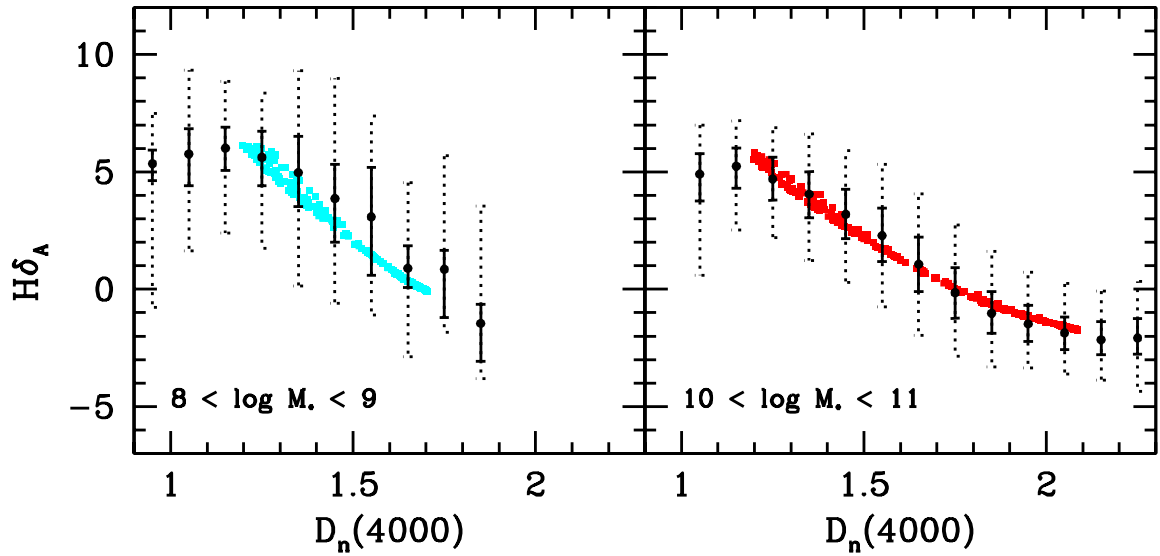}
}
\caption{\label{fig4}
\small
Black solid symbols show the median value of H$\delta_A$ as a function of D$_n$(4000) in two
different stellar mass ranges. The solid and dotted errorbars indicate the 25th to 75th
and 5th to 95th percentile ranges of the distribution of H$\delta_A$.
Coloured symbols indicate the locus occupied by galaxies with
continuous star formation histories. In the left hand panel, we have plotted
25\% solar models and in the right panel we plot solar metallicity models.}
\end {figure}
\normalsize

\begin{figure}
\centerline{
\epsfxsize=16cm \epsfbox{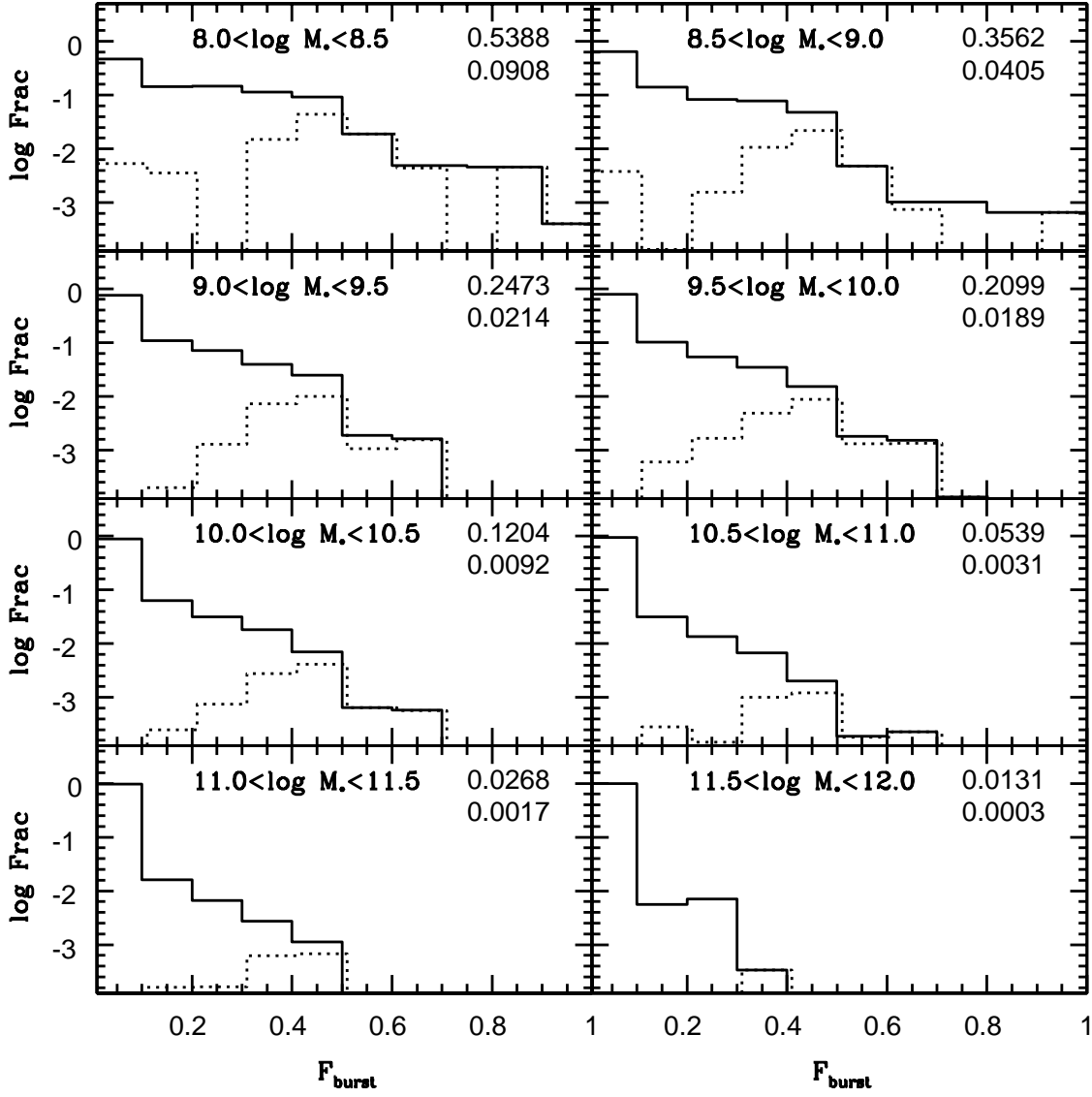}
}
\caption{\label{fig5}
\small
The solid histograms show the distributions of galaxies 
as a function of median value of the likelihood distribution
of  $F_{burst}$ in 8 stellar mass ranges.
The dotted histograms show the distributions of the median
value of  $F_{burst}$ for the high-confidence bursty galaxies in each mass
range.
 The numbers on each 
panel indicate the fraction of galaxies in each mass bin with $F_{burst}(50\%)>0$
(upper) and $F_{burst}(2.5\%)>0$ (lower).}
\end {figure}
\normalsize

\begin{figure}
\centerline{
\epsfxsize=12.4cm \epsfbox{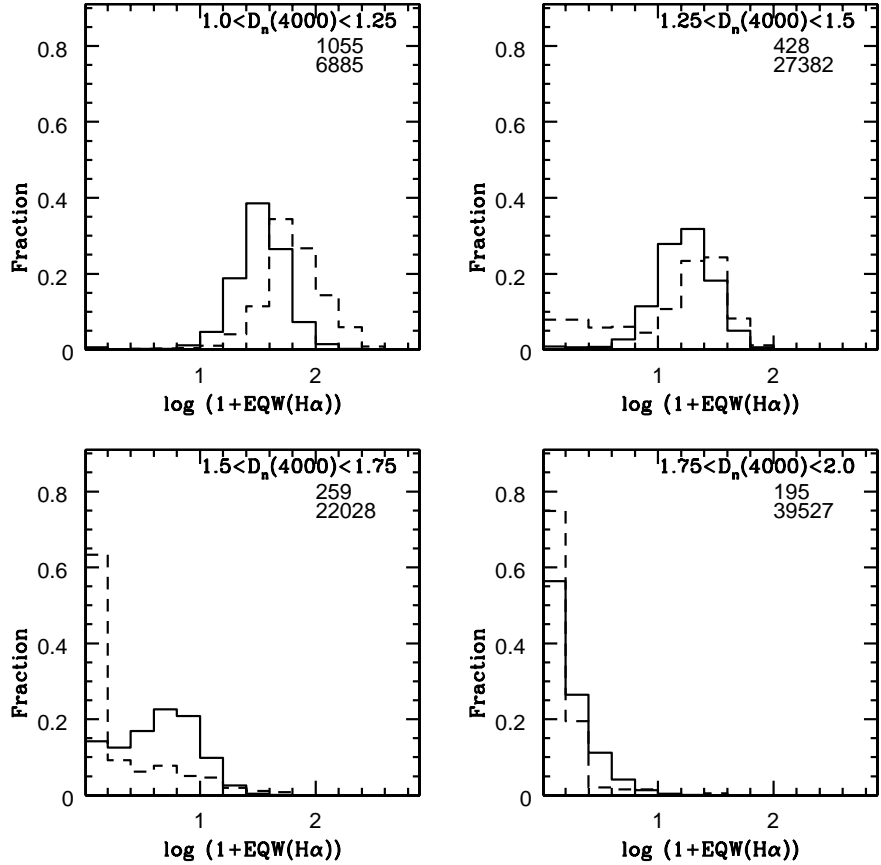}
}
\caption{\label{fig6}
\small
The distribution of H$\alpha$ equivalent widths for galaxies with 
$F_{burst}(2.5\%)>0$ (dashed))
is compared to that for galaxies with $F_{burst}(50\%)=0$ (solid) 
for four different ranges in the value of the D$_n$(4000) index.
The numbers listed in the panels indicate the number of objects
in the bursty sample (top) and in the continuous
sample (bottom).}
\end {figure}
\normalsize

\begin{figure}
\centerline{
\epsfxsize=12.4cm \epsfbox{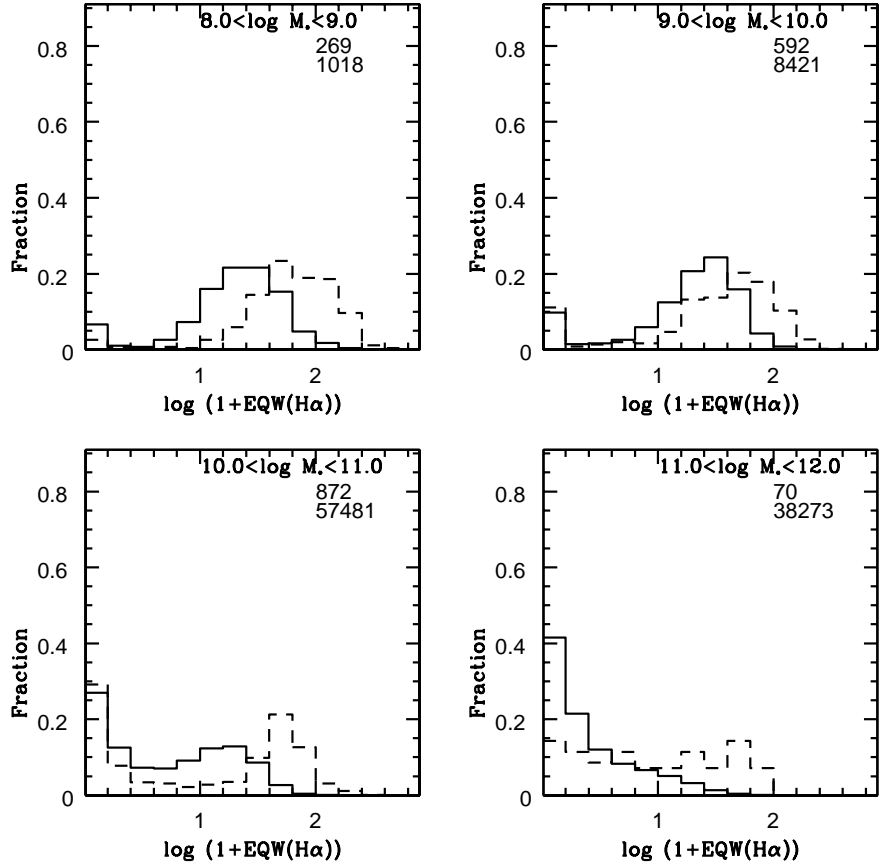}
}
\caption{\label{fig7}
\small
The distribution of H$\alpha$ equivalent widths for galaxies with 
$F_{burst}(2.5\%)>0$ (dashed))
is compared to that for galaxies with $F_{burst}(50\%)=0$ (solid) 
for four different ranges in stellar mass.                               
The numbers listed in the panels indicate the number of objects
in the bursty sample (top) and in the continuous
sample (bottom).}
\end {figure}
\normalsize

\begin{figure}
\centerline{
\epsfxsize=16cm \epsfysize=14cm \epsfbox{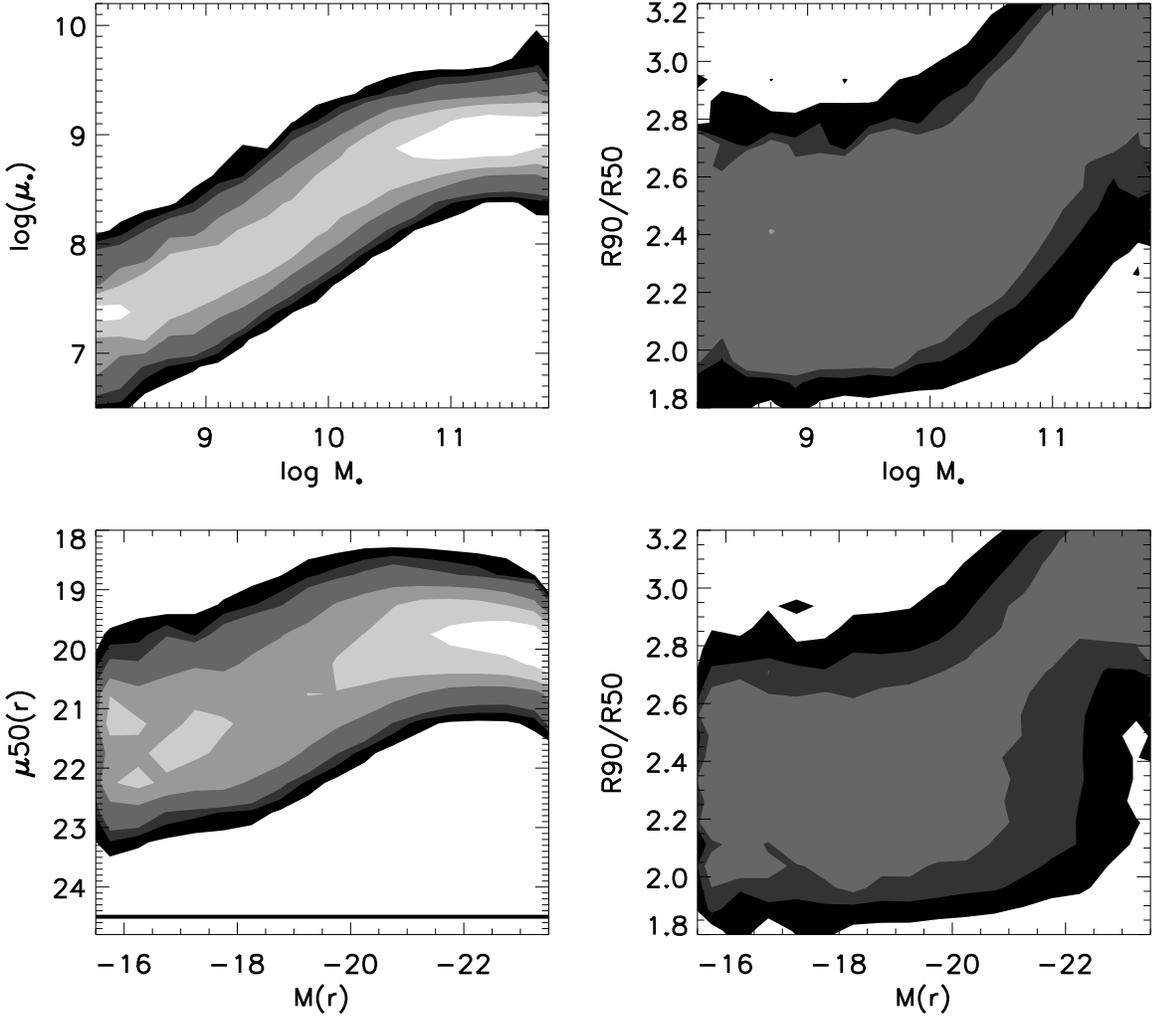}
}
\caption{\label{fig8}
\small
Conditional density distributions showing trends in the structural parameters 
$\mu_*$, $\mu_{1/2}$ and $C=R90/R50$ 
as a function the logarithm of stellar mass and as a function of $r$-band
absolute magnitude.
Galaxies have been weighted by $1/V_{max}$ and the bivariate
distribution function has been normalized to a fixed number of
galaxies in each bin of $\log M_*$ and of $r$-band   
absolute magnitude. The line in the bottom left panel indicates
the surface brightness completeness limit of the SDSS survey.}
\end {figure}
\normalsize

\begin{figure}
\centerline{
\epsfxsize=16cm  \epsfbox{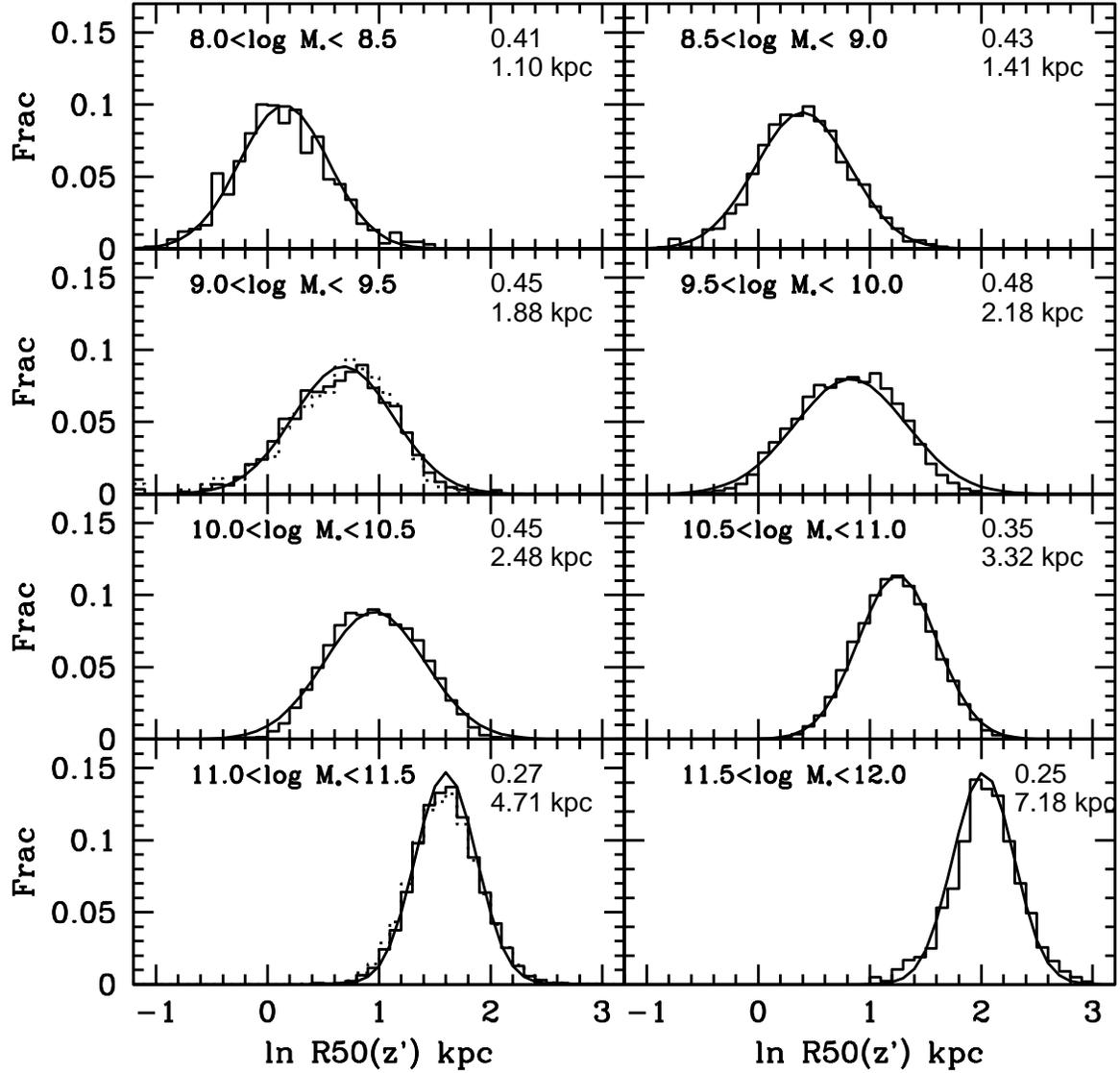}
}
\caption{\label{fig9}
\small
Solid histograms show the fraction of galaxies as a function of the 
natural logarithm
of the half-light radius in the $z$-band ($R50(z)$) in 8 different
ranges of stellar mass. The curves show lognormal fits to 
the distribution and the numbers in each panel are the best-fit 
values of $\sigma$ (top) and $R_{med}$ (bottom). The dotted histograms in
the panels with $9 < \log M_* < 9.5$ and $11 < \log M_* < 11.5$ are for
the subsample of galaxies with $V/V_{max} < 0.5$.}
\end {figure}
\normalsize

\begin{figure}
\centerline{
\epsfxsize=16cm \epsfbox{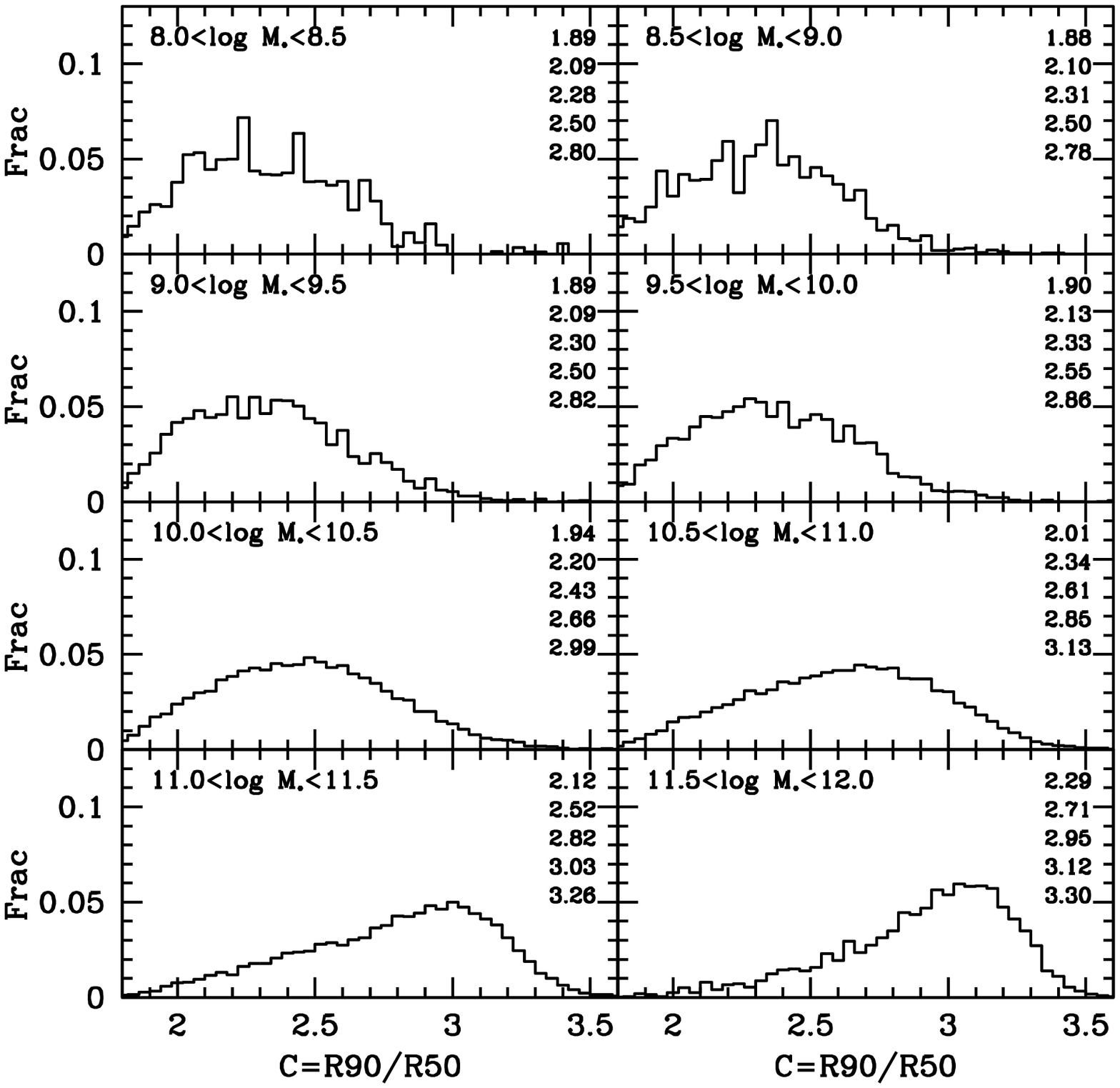}
}
\caption{\label{fig10}
\small
Histograms showing the distribution of galaxies as a function of $C$ 
index for 8 different
ranges of stellar mass. The numbers in each panel list, from top to bottom, the
5th, 25th, 50th, 75th and 95th percentiles of the distribution.}
\end {figure}
\normalsize

\begin{figure}
\centerline{
\epsfxsize=16cm \epsfbox{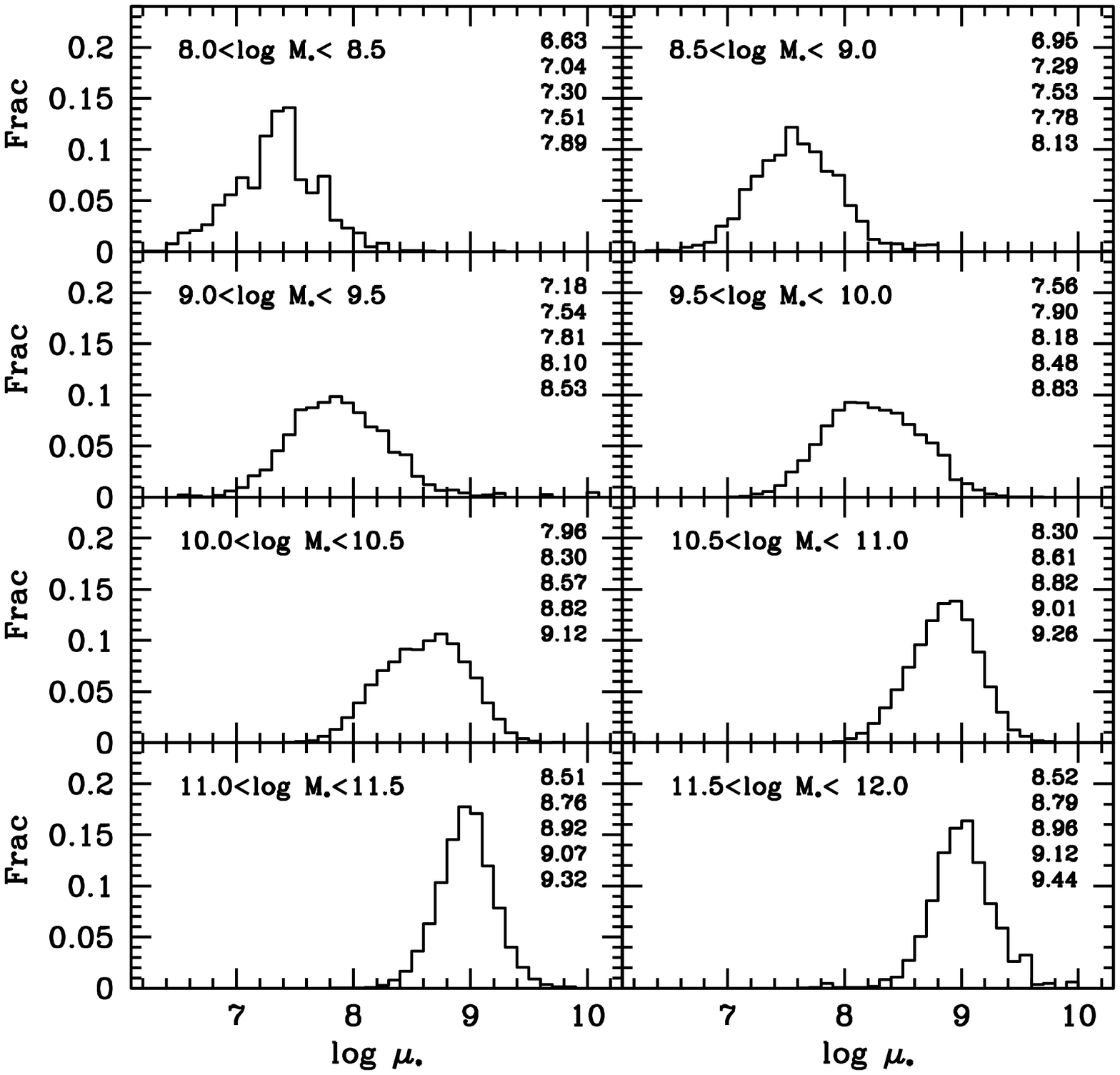}
}
\caption{\label{fig11}
\small
Histograms showing the distribution of galaxies as a function 
of $\log \mu_*$ for 8 different
ranges in stellar mass. The numbers in each panel list, from top to bottom, the
5th, 25th, 50th, 75th and 95th percentiles of each distribution.}
\end {figure}
\normalsize

\begin{figure}
\centerline{
\epsfxsize=16cm \epsfysize=14cm \epsfbox{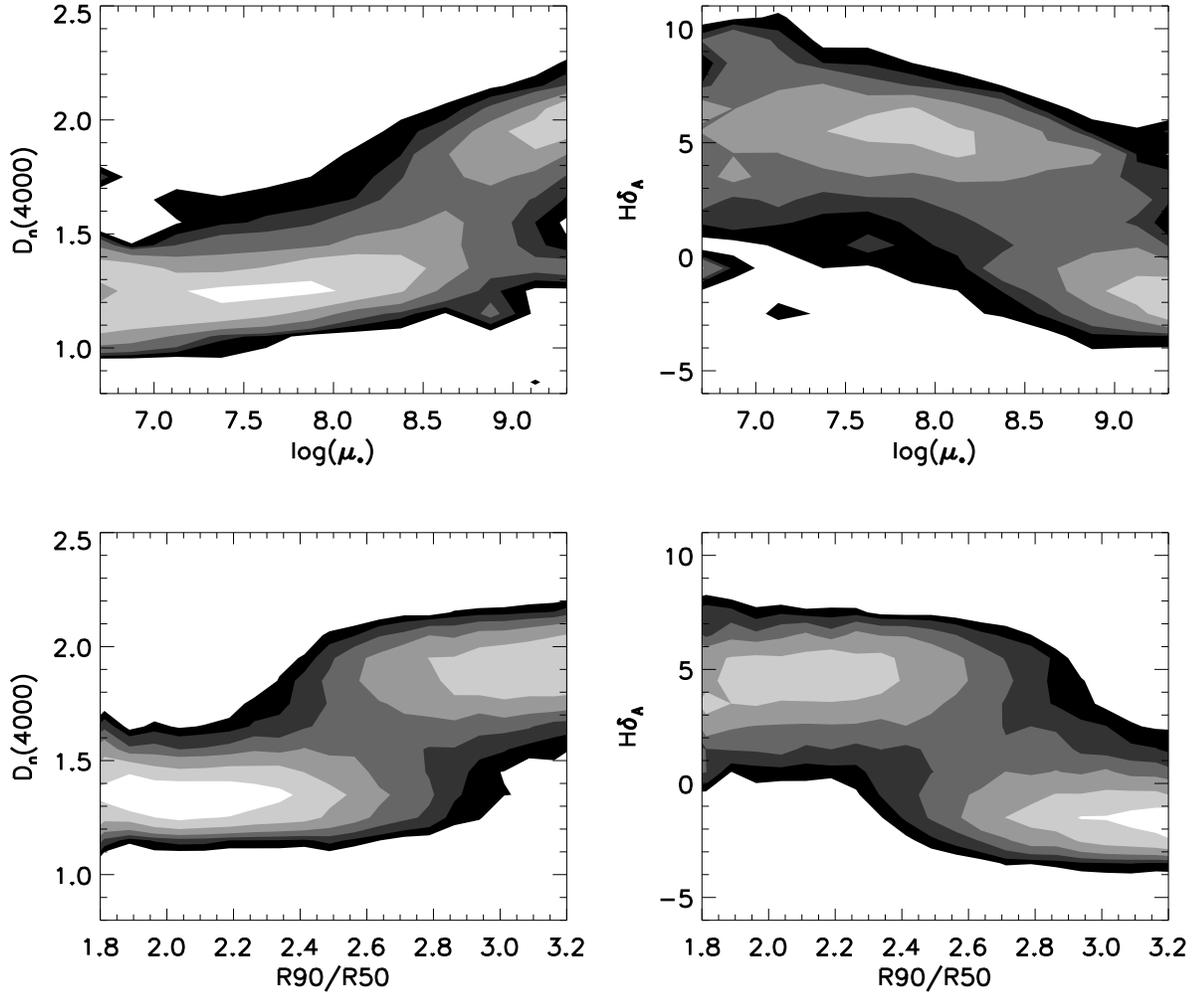}
}
\caption{\label{fig12}
\small
Conditional density distributions showing trends in the stellar age indicators
D$_n$(4000) and H$\delta_A$ 
as functions of the logarithm of the surface mass density $\mu_*$ and 
of the concentration index $C$.}
\end {figure}
\normalsize

\begin{figure}
\centerline{
\epsfxsize=14.0cm \epsfbox{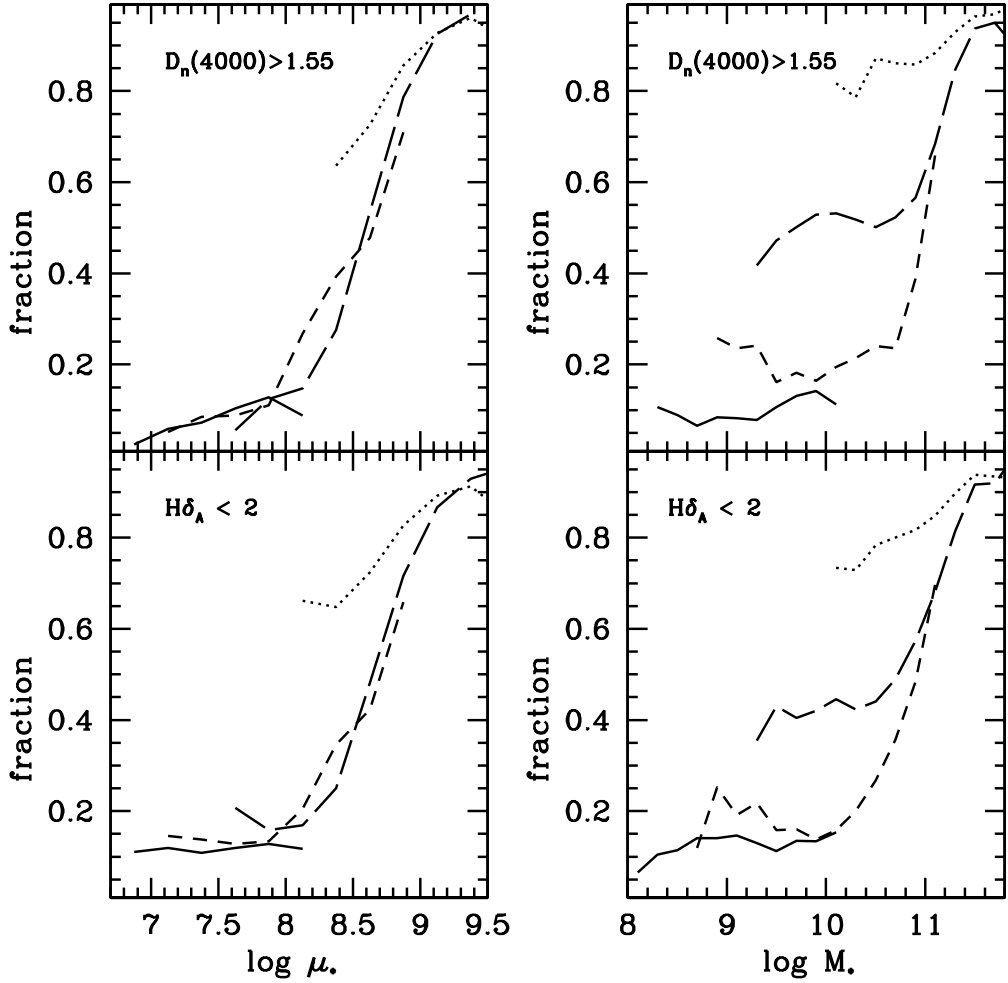}
}
\caption{\label{fig13}
\small
{\em Top Left:} The fraction of galaxies with 
 D$_n$(4000)$>1.55$ as a function of $\log \mu_*$ in fixed
ranges in stellar mass. Solid, short-dashed, long-dashed and dotted lines are for galaxies
with $\log M_*$ in the ranges 8-9, 9-10, 10-11 and 11-12, respectively.
{\em Top Right:} The fraction of galaxies with  D$_n$(4000)$>1.55$ as a function of $\log M_*$ in fixed
ranges in stellar surface mass density. Solid, short-dashed, long-dashed and dotted lines are for galaxies
with $\log \mu_*$ in the ranges 7.0-7.8, 7.8-8.3, 8.3-8.8 and 8.8-9.3, respectively.
{\em Bottom  Left:} The fraction of galaxies with 
 H$\delta_A<2$ as a function of $\log \mu_*$ in the same ranges in $M_*$. 
{\em Bottom Right:} The fraction of galaxies with  H$\delta_A< 2$ as a function of $\log M_*$ in the same
ranges in $\mu_*$.}
\end {figure}
\normalsize

\begin{figure}
\centerline{
\epsfxsize=17.4cm \epsfbox{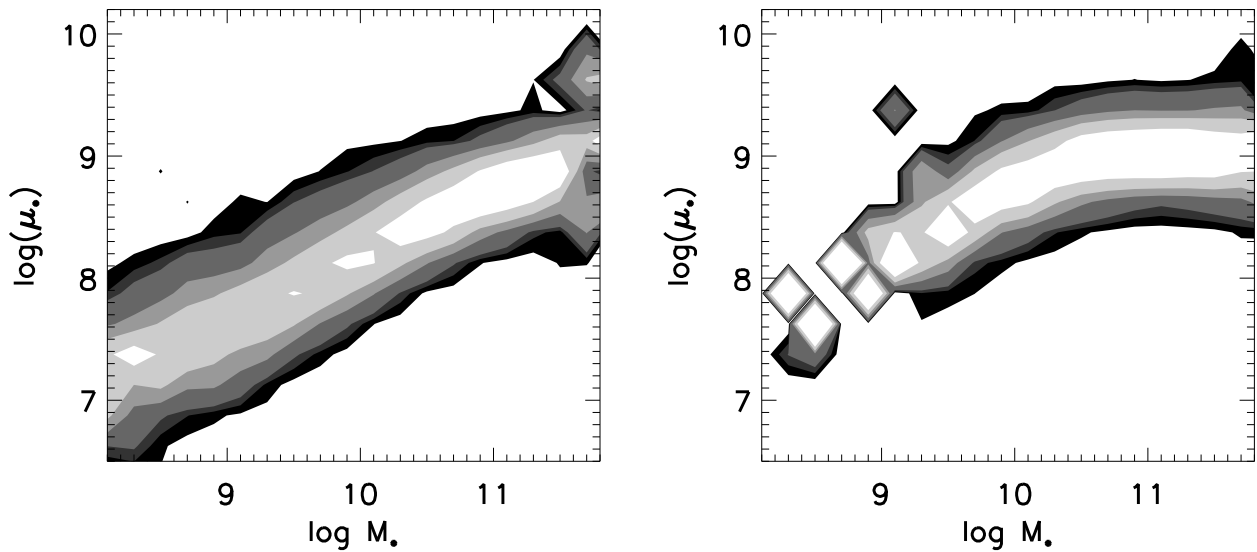}
}
\caption{\label{fig14}
\small
Conditional density distributions showing trends in
$\mu_*$ 
as a function the logarithm of stellar mass for galaxies with $D_n(4000)<1.55$
and $C<2.6$ (left) and with $D_n(4000)>1.55$ and $C>2.6$(right).}
\end {figure}
\normalsize

\pagebreak 
\Large
\begin {center} {\bf References} \\
\end {center}
\normalsize
\parindent -7mm  
\parskip 3mm

Balogh, M.L., Morris, S.L., Yee, H.K.C., Carlberg, R.G., Ellingson, E., 1999, 
ApJ, 527, 54 

Balogh, M.L., Christlein, D., Zabludoff, A.I., Zaritsky, D., 2001, ApJ, 557, 117

Barnes, J., Efstathiou, G., 1987, ApJ, 319, 575

Bell, E.F, \& de Jong, R.S., 2000, MNRAS, 312, 497  

Bell, E.F., de Jong, R.S., 2001, ApJ, 550, 212

Blanton, M.R., Dalcanton, J., Eisenstein, D., Loveday, J., Strauss, M.A., SubbaRao, M.,
Weinberg, D.H., Andersen, J.E. et al., 2001, AJ, 121, 2358

Blanton, M.R., Brinkmann, J., Csabai, I., Doi, M., Eisenstein, D., Fukugita, M., Gunn, J.E.,
Hogg, D.W., Schlegel, D.J., 2002, AJ, submitted

Bohringer, H., Belsole, E., Kennea, J., Matsuhita, K., Molendi, S., Worrall, D.M., Mushotzsky, R.F., 
Ehle, M.  et al., 2001, A\&A, 365, L181

Boselli, A., Gavazzi, G., Donas, J., Scodeggio, M., 2001, AJ, 121, 753

Bruzual, A.G., Charlot, S., 1993, ApJ, 405, 538

Bruzual, A.G., Charlot, S., 2002, in preparation

Christlein, D., 2000, ApJ, 545, 145

Cole, S., Aragon-Salamanca, A., Frenk, C.S., Navarro, J.F., Zepf, S.E., 1994, MNRAS,
271, 781

Cole, S., Lacey, C., 1996, MNRAS, 281, 716

de Blok, W.J.G., van der Hulst, J.M., Bothun, G.D., 1995, MNRAS, 274, 235

de Blok, W.J.G., McGaugh, S.S., van der Hulst, J.M., 1996, MNRAS, 283, 18

de Jong, R.S., Lacey, C., 2000, ApJ, 545, 781

Diaferio, A., Kauffmann, G., Balogh, M.L., White, S.D.M., Schade, D., Ellingson, E., 2001, MNRAS, 328, 726

Efstathiou, G., 2000, MNRAS, 317, 697

Fabian, A.C., Mushotzsky, R.F., Nulsen, P.E.J., Peterson, J.R., 2001, MNRAS, 321, L20

Fall, S.M., Efstathiou, G., 1980, MNRAS, 193, 189

Fukugita, M., Ichikawa, T., Gunn, J.E., Doi, M., Shimasaku, K., Schneider, D.P., 1996,
AJ, 111, 1748

Grebel, E.K., 2000, Proc. 33rd ESLAB Symposium, ESA SP-445, eds.F.Favata, A.A.Kaas \& Wilson 
(Noordwijk:ESA), 87-98 (astro-ph/0005296)

Gunn, J.E., Carr, M., Rockosi, C., Sekiguchi, M., Berry, K., Elms, B., de Haas, E.,
Ivezic, Z. et al, 1998, AJ, 116, 3040 

Hogg, D.W., Finkbeiner, D.P., Schlegel, D.J., Gunn, J.E., 2001, AJ, 122, 2129

Hubble, E.P., 1926, ApJ, 64, 321

Kauffmann, G., White, S.D.M., Guiderdoni, B., 1993, MNRAS, 264, 201

Kauffmann, G., Heckman, T.M., White, S.D.M., Charlot, S., Tremonti, C., Peng, E.W.,
Seibert, M., Bernardi, M.  et al. 2002, MNRAS, 
submitted (Paper I)  

Kennicutt, R.C., 1983, ApJ, 272, 54

Kennicutt, R.C., 1998, ARA\&A, 36, 189

Kroupa, P., 2001, MNRAS, 322, 231

Kuntschner, H., Lucey, J.R., Smith, R.J., Hudson, M.J., Davies, R.L., 2001, MNRAS, 323, 615

Lemson, G., Kauffmann, G., 1999, MNRAS, 302, 111

Lin, H., Kirshner, R.P., Shectman, S.A., Landy, S.D., Oemler, A., Tucker, D.L., Schechter, P.L., 1996,
ApJ, 464, 60

Loveday, J., Peterson, B.A., Efstathiou, G., Maddox, S.J., 1992, ApJ, 390,338

Marzke, R.O., Geller, M.J., Huchra, J.P., Corwin, H.G., 1994, AJ, 108, 437

Marzke, R.O., Da Costa, L.N., Pellegrini, P.S., Willmer, C.N.A., Geller, M.J., 1998, ApJ, 503,617

McKee, C.F., Ostriker, J.P., 1977, ApJ, 218, 148

Mo, H.J., Mao, S., White, S.D.M., 1998, MNRAS, 295, 319

Papovich, C., Dickinson, M., Ferguson, H.C., 2001. ApJ, 559, 620

Pettini, M., Shapley, A.E., Steidel, C.C., Cuby, J.G., Dickinson, M., Moorwood, A.F.M.,
Adelberger, K.L., Giavalisco., M., 2001, ApJ, 554, 981

Pier, J.R. et al., 2002, AJ, submitted

Roberts, M.S. \& Haynes, M.P., 1994, ARA\&A, 32, 115

Shapley, A.E., Steidel, C.C., Adelberger, K.L., Dickinson, M., Giavalisco., M., Pettini, M., 2001,
ApJ, 562, 95

Shimasaku, K., Fukugita, M., Doi, M., Hamabe, M., Ichikawa, T., Okamura, S., 
Sekiguchi, M., Yasuda, N. et al, 2001, AJ, 122, 1238

Smith, J.A., Tucker, D.L., Kent, S., Richmond, M.W., Fukugita, M.,
Ichikawa, T., Ichikawa, S.I.,Jorgensen, A.M.  et al. 2002, AJ, in press

Steidel, C.C., Adelberger, K.L., Giavalisco, M., Dicksinson, M., Pettini, M., 1999, ApJ, 519, 1

Stoughton, C., Lupton, R.H., Bernardi, M., Blanton, M.R., Burles, S., Castander, F.J.,
Connolly, A.J., Eisenstein, D.J. et al., 2002, AJ, 123, 485

Strauss, M.A., Weinberg, D.H., Lupton, R.H., Narayanan, V.K., Annis, J.,
Bernardi, M., Blanton, M., Burles, S.  et al, 2002, AJ, submitted

Strateva, I., Ivezic, Z., Knapp, G.R., Narayanan, V.K., Strauss, M.A., Gunn, J.E., 
Lupton, R.H., Schelger, D. et al, 2001, AJ, 122, 1861 

Syer, D., Mao, S., Mo, H.J., 1999, MNRAS, 305, 357

Thompson, R.I., Weymann, R.J., Storrie-Lombardi, L.J., 2001, ApJ, 546, 694

Trager, S.C., Worthey, G., Faber, S.M., Burstein, D., Gonzalez, J.J., 1998, ApJS, 116, 1

Tremonti, C.A. et al., 2003, in preparation

Warren, M.S., Quinn, P.J., Salmon, J.K., Zurek, W.H., 1992, ApJ, 399, 405

White, S.D.M., Frenk, C.S., 1991, ApJ, 379, 52

Wong, T., Blitz, L., 2002, ApJ, 569, 157

Worthey, G., Ottaviani, D.L., 1997, ApJS, 111, 377

York D.G., Adelman J., Anderson J.E., Anderson S.F., Annis J., Bahcall N.A., Bakken J.A.,
Barkhouser R. et al., 2000, AJ, 120, 1579

Zucca, E., Zamorani, G., Vettolani, G., Cappi, A., Merighi, R., Mignoli, M., Stirpe, G.M.,
MacGillivray, H. et al., 1997, A\&A, 326, 477

\end{document}